\documentclass[aps,prb,amsmath,amssymb,twocolumn]{revtex4-2}

\usepackage{graphicx}
\usepackage{adjustbox}
\usepackage{bm}
\usepackage{color}
\usepackage{braket}
\usepackage{standalone}
\usepackage{multirow}
\usepackage{tikz}
\usepackage{mathrsfs}
\usepackage[utf8]{inputenc}
\usepackage{dsfont}
\usepackage[colorlinks,bookmarks=true,citecolor=blue,linkcolor=red,urlcolor=blue]{hyperref}
\usepackage{bbm}

\definecolor{red}{rgb}{1,0,0}
\definecolor{green}{rgb}{0,1,0}
\definecolor{blue}{rgb}{0,0,1}

\newcommand{\eq}[1]{Eq.~(\ref{#1})}
\newcommand{\hc}{\text{H.c.}}

\usepackage[caption=false]{subfig}
\usepackage{bbm}
\usepackage{floatrow}
\usepackage{verbatim}
\usepackage{mathtools}

\usepackage[acronym]{glossaries}

\makenoidxglossaries

\newacronym{dmrg}{DMRG}{density matrix renormalization group}
\newacronym{caz}{CAZ}{Cartan-Altland-Zirbauer}
\newacronym{2d}{2D}{two-dimensional}
\newacronym{1d}{1D}{one-dimensional}
\newacronym{sc}{SC}{superconducting}
\newacronym{gs}{GS}{ground state}
\newacronym{mps}{MPS}{matrix product state}
\newacronym{mpste}{MPSTE}{matrix product state time evolution}
\newacronym{bs}{BS}{backscattering}
\newacronym{fs}{FS}{forward scattering}
\newacronym{pbc}{PBC}{periodic boundary conditions}
\newacronym{obc}{OBC}{open boundary conditions}
\newacronym{bch}{BCH}{Baker-Campbell-Hausdorff}
\newacronym{gsp}{GSP}{ground state preparation}

\glsdisablehyper

\begin{document}

\title{Adiabatic preparation of a number-conserving atomic Majorana phase}
\author{Benjamin Michen}
\email{benjamin.michen@tu-dresden.de}
\author{Tim Pokart}
\email{tim.pokart@tu-dresden.de}
\author{Jan Carl Budich}
\email{jan.budich@tu-dresden.de}
\affiliation{Institute of Theoretical Physics${\rm ,}$ Technische Universit\"{a}t Dresden and W\"{u}rzburg-Dresden Cluster of Excellence ct.qmat${\rm ,}$ 01062 Dresden${\rm ,}$ Germany}
\date{\today}

\begin{abstract}
We construct a protocol to adiabatically prepare the ground state of a widely discussed number-conserving model Hamiltonian for ultracold atoms in optical lattices that supports Majorana edge states. In particular, we introduce a symmetry breaking mass term that amounts to threading a commensurate (artificial) magnetic flux through the plaquettes of the considered two-leg ladder which opens a constant bulk gap. This enables the preparation of the topological Majorana phase from a trivial Mott insulator state with optimal asymptotic scaling of the ramp time in system size, which is linear owing to the critical nature of the target state. Using constructive bosonization techniques that account for both finite size effects and global fermion number conservation, we are able to fully explain with theory the somewhat counterintuitive necessity of the aforementioned commensurate flux for a controlled bulk gap. Our analytical predictions are corroborated and quantified by unbiased numerical \gls{mps} simulations. Directly building on previous experimental work, the crucial flux-term of the proposed protocol is feasible with state-of-the-art experimental techniques in atomic quantum simulators. 
\end{abstract}

\maketitle

\section{Introduction}
Quantum simulators, e.g., based on  ultracold atoms in optical lattices \cite{Ultracold_atoms_1, Ultracold_atoms_2, Ultracold_atoms_3, Ultracold_atoms_4, Ultracold_atoms_5, HH_model_real, Hall_Ribbons}, offer an impressive flexibility in engineering many-body Hamiltonians, now rendering the preparation of low-entropy states of complex quantum matter a key remaining challenge in the field . In this context, topological states inevitably separated from a trivial product state by a topological quantum phase transition are of primary interest \cite{HasanKane2010, Qi2011, Budich2013, Wen2017}. A conceptually simple example of this type is provided by topological superconductors hosting Majorana bound states \cite{Majorana_review_1, Majorana_review_2, Majorana_review_3, Majorana_review_4, Majorana_TSC_1, Majorana_TSC_2, Majorana_TSC_3, Majorana_TSC_4}. These exotic quasiparticles have been widely discussed for topological quantum computing architectures \cite{TQC_1, TQC_2, TQC_3, TQC_4, TQC_5, TQC_6, TQC_7} and related settings \cite{Majorana_FQH_1, Majorana_FQH_2, Majorana_FQH_3, Majorana_FQH_4, Majorana_FQH_5, Majorana_FQH_6, Kitaev_Honeycomb} due to their non-Abelian braiding statistics \cite{Non_abelian_statistics_1, Non_abelian_statistics_2, Non_abelian_statistics_3, Non_abelian_statistics_4}.

\begin{figure}[thp!]	 
  {
        \vbox to 0pt {
                \raggedright
                \textcolor{white}{
                    \subfloatlabel[1][Fig:illustration_a]
                    \subfloatlabel[2][Fig:illustration_b]
                    \subfloatlabel[3][Fig:illustration_c]
                }
            }
    }
\includegraphics{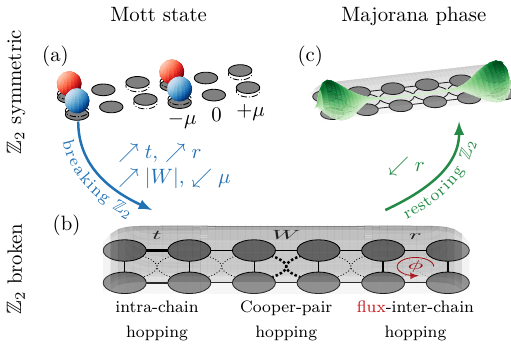}
\caption{Illustration of the protocol in  \eq{Eqn:time_dependence_p1}. (a) Starting from a Mott state  stabilized at filling $\nu = 1/3$ by a staggered chemical potential $\mu$, in (b) the pair hopping term $W$ may be adiabatically switched on thanks to an additional interchain hopping $r$ that threads a commensurate flux $\phi = \nu$ through each plaquette. Importantly, this flux-hopping breaks the $\mathbb{Z}_2$ wire-parity symmetry $P_\mathrm{a} = (-1)^{N_\mathrm{a}}$ and induces a bulk gap in combination with either of both $\mu$ and $W$. (c) Finally, the target Hamiltonian hosting the atomic Majorana phase is approached by turning off $r$.  As the final state is critical, this requires a specialized strategy to optimize preparation time (see Sec.~\ref{Sec:crit_state_prep}).}\label{Fig:illustration}
\end{figure}	

The paradigmatic example of a Majorana phase is known as the Kitaev chain \cite{Kitaev_chain, KC_related_models_1, KC_related_models_2, KC_related_models_3, KC_related_models_4}, a \gls{1d} (proximity induced \cite{1D_TSC_from_proximity_1, 1D_TSC_from_proximity_2, 1D_TSC_from_proximity_3, 1D_TSC_from_proximity_4, 1D_TSC_from_proximity_5, 1D_TSC_from_BEC_proximity_1, 1D_TSC_from_BEC_proximity_2}) superconductor with global fermion parity conservation guaranteed by the bulk superconducting gap, and a single zero-energy Majorana bound state at each end. Remarkably, a variant of the Kitaev chain with global particle number conservation has been tailored for the toolbox of ultracold atoms in optical lattices  \cite{Tu_et_al, Fidkowski_et_al, Halperin_Sarma, Number_conserving_Majorana_1, Number_conserving_Majorana_2, Number_conserving_Majorana_3, Number_conserving_Majorana_4, Number_conserving_Majorana_5, Number_conserving_Majorana_6, Number_conserving_Majorana_7, Zoller_model, Goldman_model}. There, a \gls{1d} two-leg ladder (or double wire) system (see Fig. \ref{Fig:illustration}), in which inter-chain pair hopping provides the counterpart of proximity induced Cooper pair tunneling, has been shown to also stabilize a single Majorana end state. However, the absence of the bulk superconductor has two important consequences. First, the resulting Majorana phase is symmetry protected by a sub-wire fermion parity $P_\mathrm{a} = (-1)^{N_\mathrm{a}}$ with the particle number $N_\mathrm{a}$ in wire $\mathrm{a}$ \cite{Tu_et_al, Fidkowski_et_al, Halperin_Sarma}. Since $P_\mathrm{a}$ may be broken by simple single particle inter-wire tunneling, the atomic Majorana phase becomes a conventional symmetry protected topological phase. Second, the closed 1D nature of the double wire system limits the counterpart of superconducting order to the emergence of a power-law decaying pair-hopping induced gap \cite{Tu_et_al, Fidkowski_et_al, Halperin_Sarma, Mermin_Wagner, Hohenberg, SC_in_1D}. From a vantage point of state preparation, the first point is good news while the second one represents an extra challenge. More specifically, symmetry protection can be exploited by controlled intermediate symmetry breaking and restoring during the protocol, while the smaller gap of the target state requires an adiabatic time-scale that grows with system size, even in the sophisticated framework of critical state preparation  \cite{crit_state_prep_1, crit_state_prep_2, crit_state_prep_3, crit_state_prep_4, Review_QPT_Dziarmaga}. Yet, the simplest conceivable symmetry breaking term in the form of a single-particle inter-chain hopping with real strength $r$ (cf. $\phi = 0$ case below) has been found not to open a bulk gap \cite{Zoller_model,Tu_et_al, Fidkowski_et_al}. An efficient state preparation protocol for the atomic Majorana phase from a trivial initial state has so far remained elusive.

Here, by adding a Peierls phase that amounts to a commensurate synthetic flux to the inter-chain hopping, we identify a symmetry breaking mass term opening a constant gap in system size. This allows us to extend the notion of critical state preparation  to a protocol for preparing the number-conserving atomic Majorana phase with an optimal scaling in system size (see Fig. \ref{Fig:illustration} for an illustration). Specifically, we provide clear evidence that the target state can be prepared in a time that asymptotically scales linear in system size, while a protocol using plain finite size effects would require quadratic ramp times \cite{Adiabatic_theorem_Born_Fock, Adiabatic_theorem_Kato}. Our in-depth theoretical analysis of the crucial single-particle flux-hopping term is based on analytical techniques in the framework of constructive Bosonization \cite{Bosonization_OBC, Bosonization_Delft, Bosonization_Schoenhammer, Giamarchi} that take into account finite size terms and resolve effects sensitive to the total fermion parity. The resulting qualitative predictions are then corroborated and quantified by numerical simulations using \gls{mps} methods \cite{Itensor_basic, Itensor_codebase_release}. 

We start by introducing the model Hamiltonian and the considered perturbations in Sec.~\ref{Sec:Model}, before we proceed to describe and investigate the \gls{gs} preparation protocol in detail in Sec.~\ref{Sec:GSP}. Our bosonization analysis is presented in Sec.~\ref{Sec:Bosonization}, and a concluding discussion in \ref{Sec:conclusion}.

\section{Model} \label{Sec:Model}
The starting point of our analysis is the model proposed in Ref.~\cite{Zoller_model}, the ground state of which is the target state of our present preparation protocol. It is given by a free Hamiltonian $H_0$ consisting of two quantum chains with a nearest-neighbor hopping $t$ and a pair hopping term $H_W$, which read
\begin{align}
H_0 &= -t\sum_{\gamma = \mathrm{a,b}}\sum_{j = 1}^{L-1} \left[(c_{\gamma,j}^\dagger c_{\gamma,j+1} + c_{\gamma,j+1}^\dagger c_{\gamma,j}) \right], \nonumber \\ 
H_W &= W \sum_{j = 1}^{L-1} \left[c_{\mathrm{a},j}^\dagger c_{\mathrm{a},j+1}^\dagger c_{\mathrm{b},j} c_{\mathrm{b},j+1} + \text{H.c.} \right], \label{Eqn:Zoller_model}
\end{align}
where $c_{\gamma,j}$ annihilates a fermion at site $j$ of wire $\gamma = \mathrm{a,b}$.
An experimental implementation with ultracold atoms in optical lattices is discussed in detail in the original publication \cite{Zoller_model}. Energy is measured in units of the hopping amplitude such that $t = 1$. Unless stated otherwise, we set $W = -1.8$, use \gls{obc}, and work at a fixed (even) particle number $N_\mathrm{tot} = N_\mathrm{a} + N_\mathrm{b}$ such that the filling fraction is $\nu = \frac{1}{3}$. For these parameters, a Majorana phase protected by the wire parity $P_\mathrm{a} = (-1)^{N_\mathrm{a}}$ emerges \cite{Tu_et_al, Fidkowski_et_al, Halperin_Sarma, Zoller_model}, see also Fig.~\ref{Fig:illustration_c}.

To facilitate the adiabatic transition between the Mott state and the Majorana phase, we introduce two perturbations in the form of an interchain tunneling $H_\phi$ and a staggered chemical potential $H_\mu$, reading
\begin{align}
H_\phi &= r \sum_{j = 1}^L  \left[ e^{2 \pi i \phi j}c_{\mathrm{a},j}^\dagger c_{\mathrm{b},j} + e^{-2 \pi i \phi j}c_{\mathrm{b},j}^\dagger c_{\mathrm{a},j} \right], \label{Eqn:H_phi} \\
H_\mu &= \mu \sum_{\gamma = \mathrm{a, b}} \sum_{j = 1}^L \left[n_{\gamma, 3j -2}  - n_{\gamma, 3j}\right ], \label{Eqn:H_mu}
\end{align}
where all parameters are real. $H_\phi$ threads an artificial magnetic flux $\phi$ through each plaquette of the ladder and breaks the protecting $\mathbb Z_2$ symmetry for any value of $\phi$. However, as we derive later on from the bosonization picture (see Sec.~\ref{Sec:Bosonization}), interchain \gls{bs} is necessary to drive the system away from the critical point and induce a stable topologically trivial phase. This requires a flux $\phi$ that is commensurate with the filling fraction, hence we choose $\phi = \nu = \frac{1}{3}$ in the following unless stated otherwise. The role of the staggered chemical potential $H_\mu$ will be to introduce a single particle gap at filling $\nu = \frac{1}{3}$ and stabilize the Mott state (cf. Fig.~\ref{Fig:illustration_a}). Both perturbation terms are well within reach of current experimental techniques, most notably precise control over artificial fluxes has been demonstrated by photon-assisted tunneling \cite{Ultracold_atoms_1, Ultracold_atoms_2, Ultracold_atoms_5, Hall_Ribbons}.

\section{Ground state preparation} \label{Sec:GSP}
As the excitation gap above the target state scales as $\propto 1/L$ with system size  \cite{Tu_et_al, Halperin_Sarma, Fidkowski_et_al, Zoller_model}, it resembles a critical state. Hence, we find it useful to look to the theory of theory of critical ground state preparation for guidance, where experimental control over a ``mass'' term that gaps out the critical target system is a crucial ingredient \cite{crit_state_prep_1, crit_state_prep_2, crit_state_prep_3, Review_QPT_Dziarmaga, crit_state_prep_4}. For the present model, this role is assumed by the flux-hopping with amplitude $r$. In Sec.~(\ref{Sec:introduction_protocols}), we introduce a protocol that exploits this term to optimize \gls{gs} preparation time by considering the total Hamiltonian $H_0(\tau) + H_W(\tau) + H_\phi(\tau) + H_\mu(\tau)$ as per Eqs.~(\ref{Eqn:Zoller_model}-\ref{Eqn:H_mu}) with parameters varying as a function of the dimensionless parameter $\tau$, and a second, simpler protocol that never breaks the symmetry and requires a ramp along a critical region in parameter space. To demonstrate the expected advantage of the symmetry-breaking protocol over the symmetry-respecting protocol, we conduct \gls{mpste} simulations in Sec.~\ref{Sec:crit_state_prep}.

\begin{figure}[htp!]	 
    {
        \vbox to 0pt {
                \raggedright
                \textcolor{white}{
                    \subfloatlabel[1][Fig:protocol_data_a]
                    \subfloatlabel[2][Fig:protocol_data_b]
                    \subfloatlabel[3][Fig:protocol_data_c]
                    \subfloatlabel[4][Fig:protocol_data_d]
                }
            }
    }
    \includegraphics{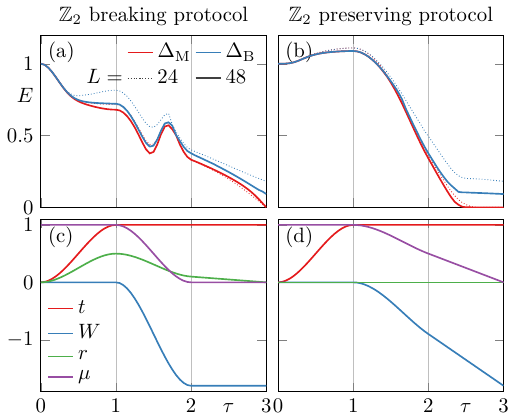}
    \caption{(a) The upper panel shows the gaps $\Delta_\mathrm{M} = E_1 - E_0$ and $\Delta_\mathrm{B} = E_2 - E_0$ as a function of $\tau$. We present data for system size $L = 24$ and $L =48$ at filling $\nu = 1/3$, corresponding to $N_\mathrm{tot} = 16$ and $N_\mathrm{tot} = 32$ particles, both with \gls{obc}. The interchain hopping phase is set to $\phi = 1/3$ while the other parameters follow the function \eq{Eqn:time_dependence_p1}, which is illustrated in the lower panel. (b) Similar to (a), but with the parameters determined by \eq{Eqn:time_dependence_p2} as illustrated in the lower panel.}
\end{figure}

\subsection{Symmetry breaking versus symmetric protocol}\label{Sec:introduction_protocols}
Using the indicator function $\mathbbm{1}_I(\tau)$ for the interval $I$ and the auxiliary functions $f(\tau) = 3\tau^2 - 2 \tau^3$, $g(\tau) = 11\tau^2 -7 \tau^3$, and $h(\tau) = 2\tau^2 - \tau^3$, the parameter functions for the symmetry-breaking protocol can be written as

\begin{align}
t_1(\tau) &= t \left [ f(\tau) \mathbbm{1}_{(0,1]}(\tau)+ \mathbbm{1}_{(1,3]}(\tau) \right ], \nonumber \\
W_1(\tau) &= W \left[f(\tau - 1) \mathbbm{1}_{(1,2]}(\tau) + \mathbbm{1}_{(2,3]}(\tau) \right], \nonumber \\
r_1(\tau) &= r \left[ 5 f(\tau) \mathbbm{1}_{(0,1]}(\tau) + [1 - g(\tau - 1)] \mathbbm{1}_{(1,2]}(\tau) \right . \nonumber \\ 
&\quad + \left. (3 - \tau) \mathbbm{1}_{(2,3]}(\tau)\right], \nonumber \\
\mu_1(\tau) &= \mu\left[\mathbbm{1}_{[0,1]}(\tau) + [1 - f(\tau-1)] \mathbbm{1}_{(1, 2]}(\tau) \right], \label{Eqn:time_dependence_p1}
\end{align}
where $t = 1$, $W = -1.8$, $r = 0.1$ and $\mu = 1$. This exemplary choice of parameter functions is motivated by simplicity, so as to illustrate a smooth path in parameter space from the trivial initial state to the target state that is fully gapped baring the critical end point. The protocol is divided into three stages corresponding to $\tau \in [0,1]$, $\tau \in (1,2]$, and $\tau \in (2,3]$; an illustration of the parameter evolution can be found in Fig.~\ref{Fig:protocol_data_c}. At $\tau = 0$, only a staggered chemical potential is present, which stabilizes a Mott state at filling $\nu = 1/3$ (cf. Fig.~\ref{Fig:illustration_a}). During the first stage, the Mott state is deformed adiabatically into a non-interacting double-leg flux ladder state, with the staggered chemical potential $\mu$ always sustaining a large gap. Stage two exchanges the chemical potential with the pair-hopping $W$ and ramps down the flux-hopping amplitude to $r = 0.1$, ending in the target Hamiltonian plus the residual flux hopping (cf. Fig.~\ref{Fig:illustration_b}). The final step consists of turning the flux hopping off entirely, thereby arriving at the target Hamiltonian hosting the Majorana phase (cf. Fig.~\ref{Fig:illustration_c}).

Fig.~\ref{Fig:protocol_data_a} shows the evolution of the gaps $\Delta_\mathrm{M} = E_1 - E_0$ and $\Delta_\mathrm{B} = E_2 - E_0$ to the first and second excited state as a function of $\tau$ for two system sizes $L = 24$ and $L=48$ in an open geometry. The data is obtained from \gls{dmrg} simulations \cite{Itensor_basic, Itensor_codebase_release}. Here, $\Delta_\mathrm{M}$ and $\Delta_\mathrm{B}$ can be interpreted as the gap to the Majorana mode and the excited bulk states, respectively. In agreement with the theoretical expectation, the Majorana gap $\Delta_\mathrm{M}$ of the target Hamiltonian decays exponentially with system size, while the bulk gap $\Delta_\mathrm{B}$ closes $\propto 1/L$ \cite{Tu_et_al, Halperin_Sarma, Fidkowski_et_al, Zoller_model}. This poses a challenge for the third stage: while stage one and two can be completed in a finite time regardless of system size due to the finite gap, the criticality of the target state requires a specialized strategy to optimize preparation time, which will be subject of the following section.

We contrast the protocol in \eq{Eqn:time_dependence_p1} with a symmetry-respecting path in parameter space designated by the parameter functions
\begin{align}
t_2(\tau) &= t_1(\tau), \nonumber \\
W_2(\tau) &= 0.5 W \left[h(\tau - 1) \mathbbm{1}_{(1,2]}(\tau) + (-1 + \tau) \mathbbm{1}_{(1,2]}(\tau)  \right ], \nonumber \\
r_2(\tau) &= 0, \nonumber \\
\mu_2(\tau) &= 0.5 \mu \left[ 2 \mathbbm{1}_{[0,1]}(\tau) + [2 - h(\tau - 1)] \mathbbm{1}_{(1, 2]}(\tau) \right . \nonumber \\ 
 &\quad+ \left. (3 - \tau) \mathbbm{1}_{(2, 3]}(\tau) \right].  \label{Eqn:time_dependence_p2}
\end{align}
Again, we illustrate the parameter evolution in Fig.~\ref{Fig:protocol_data_d} and provide \gls{dmrg} data on the evolution of the energy gaps in Fig.~\ref{Fig:protocol_data_b}. Similar to the first protocol \eq{Eqn:time_dependence_p1}, stages one and two are gapped and thus completable in finite time. The third stage not only ends in a critical state, but requires ramping along a critical line in parameter space, which is a generic problem of any parameter path that respects the protecting $\mathbb{Z}_2$ symmetry. As we demonstrate in the following section, this will severely increase the necessary preparation time.

We stress that due to the $\mathbb Z_2$ symmetry breaking, the protocol \eq{Eqn:time_dependence_p1} will prepare a superposition of the two degenerate $P_\mathrm{a}$ eigenstates for \gls{obc}, while the symmetry-respecting protocol \eq{Eqn:time_dependence_p2} prepares the state that corresponds to the parity of the initial Mott state. To prepare a definite-parity eigenstate with the first protocol, one can prepare a flux-gapped state with \gls{pbc} on the single-particle Hamiltonian $H_0$ first and then adiabatically cut the link connecting the boundaries together with the flux-hopping $r$, see also \cite{Cutting_Majorana_Chain}. We discuss this further in Appendix \ref{App:Sec:crit_state_prep} and provide \gls{mpste} data for system size $L = 24$. There, we also characterize single-particle correlations in the prepared states.

The gap formation due to $H_\phi$ exploited in the \gls{gs} preparation protocol is robust against disorder both on the hopping amplitude and phase; numerical data and a discussion are provided in Appendix \ref{App:Sec:stability_perturbations}.

\subsection{Critical state preparation}\label{Sec:crit_state_prep}
The time $\tau_\mathrm{tot}$ to adiabatically prepare a critical state is generally bounded from below by the inverse excitation gap $\Delta_\mathrm{c}^{-1}$ above the target state \cite{crit_state_prep_1, crit_state_prep_2}, which implies a minimal asymptotic scaling with system size  of $\tau_\mathrm{tot} \propto 1/L$ for the model at hand. To achieve an optimal fidelity for a given preparation time, we approach the critical point by ramping down the flux hopping as 
\begin{align}
r_p(\tau) = r_{0} |1 -  \tau / \tau_\mathrm{tot}|^p \label{Eqn:power_law_ramp}
\end{align}
for $\tau \in [0, \tau_\mathrm{tot}]$. At $r_{0} = 0.1$, this completes stage three of the protocol Fig.~\ref{Fig:protocol_data_a}, with the power $p$ controlling the transition rate close to the critical point. Based on estimates deriving from the Kibble-Zurek mechanism \cite{crit_state_prep_3, Review_QPT_Dziarmaga}, this ramp can be expected to be adiabatic if $\tau_\mathrm{tot} \gg L^{z_\mathrm{c} + \frac{1}{p \nu_\mathrm{c}}}$, where $L$ is the system size and $z_\mathrm{c}, \nu_\mathrm{c}$ are the dynamical and correlation length exponents characterizing the critical point. The critical behavior of \eq{Eqn:H_0_bosonized} stems from the symmetric sector, which is simply a free Luttinger liquid exhibiting a linear dispersion and thus $z_\mathrm{c} = 1$ and $\nu_\mathrm{c} = 1$, suggesting a quadratic scaling of preparation time for $p = 1$ and a linear scaling in the limit $p \to \infty$. However, this limit is not directly viable while keeping the initial value $r_{0}$ fixed, since the rate $|\dot{r}_p(\tau = 0)| = r_{0} p / \tau_\mathrm{tot}$ is not bounded. Nevertheless, virtually linear scaling may still be achieved if we consider increasing the power $p(L)$ sublinearly as a function of system size together with a linear scaling of the total ramp time $\tau_\mathrm{tot}(L) \propto L$ such that $\lim_{L\to \infty} p(L) / \tau_\mathrm{tot}(L) = 0$ while $\lim_{L \to \infty}p(L) = \infty$. Then, the bound $\tau_\mathrm{tot}(L) \gg L^{1 + \frac{1}{p(L)}} \approx L$ can be asymptotically satisfied while keeping the rate of change bounded, see Appendix \ref{App:Sec:crit_state_prep} for an extended discussion.

To validate this hypothesis, we consider the Hamiltonian $H_1(\tau)$ at fixed parameters $t = 1$, $W = -1.8$, $\mu = 0$, $\phi = 1/3$, and the only time-dependence $r(\tau) = r_p(\tau)$ given by \eq{Eqn:power_law_ramp} with $r_{0} = 0.1$, such that $H_1(0)$ is precisely the Hamiltonian at the beginning of stage three and $H_1(\tau_\mathrm{tot})$ is the target Hamiltonian. We time-evolve the initial ground state $\ket{\psi_\mathrm{prep}(\tau = 0)}$  of $H_1(\tau = 0)$ under $H_1(\tau)$ using  \gls{mpste} and measure the overlap of the prepared state with the two degenerate \gls{gs} $\ket{\mathrm{GS}_{1,2}}$ of $H_1(\tau_\mathrm{tot})$ as $F = |\braket{\mathrm{GS}_{1}|\psi_\mathrm{prep}(\tau_\mathrm{tot})}|^2 + |\braket{\mathrm{GS}_{2}|\psi_\mathrm{prep}(\tau_\mathrm{tot})}|^2$. The \gls{mpste} results for four system sizes ranging from $L = 24$ to $L = 96$ are depicted in Fig.~\ref{Fig:fidelity_data_a} over a time axis linearly rescaled with system size for power $p = 1$, corresponding to a linear ramp, and a higher power $p(L)$ that we optimized to find the best possible fidelity. We note that the lines do not collapse exactly at large $\tau_\mathrm{tot}$, which we attribute to a combination of subleading corrections to the expected linear scaling and limitations of the \gls{mpste} method for large system sizes. However, the general result is compatible with our hypothesis: increasing the power $p$ sublinearly with system size allows for the preparation of a ground state with high fidelity in a time that scales asymptotically linear with system size.

We perform similar simulations for the third stage of \eq{Eqn:time_dependence_p2} by considering the Hamiltonian $H_2(\tau)$ with fixed parameters $t = 1$, $r = 0$, and the time-dependence $W(\tau) = W_2(2 + \tau/\tau_\mathrm{tot})$, $\mu(\tau) = \mu_2(2 + \tau/\tau_\mathrm{tot})$  (cf. \eq{Eqn:time_dependence_p2}). This is simply a linear ramp towards the critical target state, similar to \eq{Eqn:power_law_ramp} with $p = 1$. Contrary to the symmetry-breaking protocol, this approach follows  a critical line in parameter space, hence there is nothing to gain by a larger power as the finite-size gap persists along a finite interval in time. The adiabatic theorem implies a scaling of preparation time as $\tau_\mathrm{tot}\propto \Delta_\mathrm{c}^{-2} \propto L^2$ \cite{Adiabatic_theorem_Born_Fock, Adiabatic_theorem_Kato}, suggesting that this approach severely underperforms the symmetry-breaking protocol. These expectations are confirmed by the \gls{mpste} results we present in Fig.~\ref{Fig:fidelity_data_b} for system sizes $L = 24$, $L = 48$, and $L = 72$. For $L = 24$, the time to reach $99.9\%$ fidelity is roughly twice as much as for the symmetry-breaking protocol, but for larger system sizes this worsens rapidly. We also note that while the $p = 1$ case presented in Fig.~\ref{Fig:fidelity_data_a} still greatly outperforms the protocol in Fig.~\ref{Fig:fidelity_data_b} on a quantitative level, the shape of the curves is very similar in consistency with the expected quadratic scaling of preparation time in both cases.

\begin{figure}[htp!]	 
{
        \vbox to 0pt {
                \raggedright
                \textcolor{white}{
                    \subfloatlabel[1][Fig:fidelity_data_a]
                    \subfloatlabel[2][Fig:fidelity_data_b]
                }
            }
    }
    \includegraphics{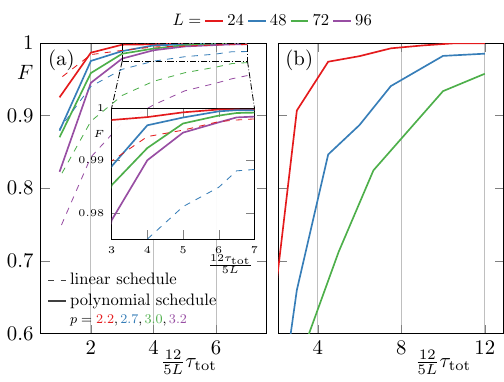}
\caption{Ground state fidelity $F$ after completing stage three of the protocol for different system sizes at filling $\nu = 1/3$. The horizontal axis indicates the total preparation time $\tau_\mathrm{tot}$ and is rescaled proportional to system size by a factor of $12 / (5 L)$. (a) Result for stage three of the symmetry-breaking protocol \eq{Eqn:time_dependence_p1} with $r(\tau)$ following \eq{Eqn:power_law_ramp} for different powers $p$. System sizes are $L = 24, 48,72,96$. (b) Result for stage three of the symmetry-preserving protocol \eq{Eqn:time_dependence_p2}, here with the parameters simply following a linear ramp over a time $\tau_\mathrm{tot}$. System sizes are $L = 24, 48,72$.}\label{Fig:fidelity_data}
\end{figure}

For completeness, we note that there exist proposals to achieve a linear scaling of critical state preparation time by a spatially inhomogeneous ramp of the ``mass'' term, where the critical region spreads through the system as a front that propagates at a well-chosen speed \cite{crit_state_prep_3, crit_state_prep_4}. We implement such a protocol numerically and provide \gls{mpste} data in Appendix \ref{App:Sec:IH_ramp}, demonstrating that this approach is clearly outperformed by the power law optimization strategy we presented above. Furthermore, for an experimental implementation the precise manipulation of the transverse hopping amplitude in space would rise as another significant challenge.

\section{Bosonization} \label{Sec:Bosonization}
We now investigate the flux-induced GS splitting in a bosonization framework tailored to \gls{obc} \cite{Bosonization_OBC}. To this end, we interpret the lattice model as the discretization of a continuous theory defined on the interval $[0, \tilde{L}]$, where $\tilde{L} = (L + 1) a_0$ with the lattice constant $a_0$. The continuum fields obey \gls{obc} $\psi_\gamma(0) = \psi_\gamma(\tilde{L}) = 0$ and the lattice operators are taken to be field operators evaluated at discrete positions $c_{\gamma, j} = \sqrt{a_0} \psi_\gamma(j a_0)$ \cite{Giamarchi}. The field operators can be decomposed into a left- and right-moving part as
\begin{align}
    \psi_\gamma(x) = e^{i k_\mathrm{F} x} \psi_{\gamma, \mathrm{R}}(x) + e^{-i k_\mathrm{F} x} \psi_{\gamma, \mathrm{L}}(x),
\end{align}
where $k_\mathrm{F}$ denotes the Fermi vector. Due to \gls{obc}, the left-movers are related to the right movers by $
    \psi_{\gamma, \mathrm{L}}(x) = -\psi_{\gamma, \mathrm{R}}(-x)$, with the field $\psi_{\gamma, \mathrm{R}}(x)$ being defined on the interval $[-\tilde{L}, \tilde{L}]$ with \gls{pbc} \cite{Bosonization_OBC}. It is thus sufficient to employ a regular bosonization identity for the right movers alone. In doing so, we follow a constructive bosonization approach \cite{Halperin_Sarma, Bosonization_Delft, Bosonization_Schoenhammer} that yields the identity
\begin{align}
    \psi_\mathrm{a, R}(x) &=  \frac{1}{\sqrt{2 \pi \alpha}} e^{i \hat k_\mathrm{a}} e^{i \frac{\pi}{\tilde{L}} \Delta N_\mathrm{a} x} e^{i [\theta_\mathrm{a}(x) + \phi_\mathrm{a}(x)]}, \nonumber                                            \\
    \psi_\mathrm{b, R}(x) &=  \frac{(-1)^{N_\mathrm{a}}}{\sqrt{2 \pi \alpha}} e^{i \hat k_\mathrm{b}} e^{i \frac{\pi}{\tilde{L}} \Delta  N_\mathrm{b} x} e^{i [\theta_\mathrm{b}(x) + \phi_\mathrm{b}(x)]}. \label{Eqn:Bosonization_identity}
\end{align}
In the above equation, $\alpha$ plays the role of a regularization parameter, $\Delta  N_\gamma = N_\gamma - \tilde{L} k_\mathrm{F} / \pi$ counts the number of $\gamma = \mathrm{a, b}$ particles relative to the Fermi surface, and the Hermitian operator $\hat k_\gamma$ is conjugate to the particle number in the sense that $[N_\gamma, e^{\pm i \hat k_{\gamma'}}] = \mp e^{\pm i \hat k_\gamma}\delta_{\gamma, \gamma'}$. Thus, the terms $e^{\pm i \hat k_\gamma}$ will decrease / increase the number of $\gamma$ particles by one and act as Klein factors together with the parity $(-1)^{N_\mathrm{a}}$. The fields $\theta_\gamma$, $\phi_\gamma$ are the usual phase fields constructed from the Fourier components of the electron density.

Using this identity, we will proceed to bosonize \eq{Eqn:Zoller_model} and \eq{Eqn:H_phi}, which is in principle straightforward albeit a bit tedious. Consequently, we only present the resulting bosonized theory here and compare its predictions to DMRG results. A detailed derivation can be found in Appendix \ref{App:Sec:Bosonization}.

\subsection{Bosonization of Majorana Hamiltonian}
After performing a canonical transformation to antisymmetric/symmetric combinations of the fields
\begin{align}
    \hat \theta_{\pm}(x) &=  \frac{1}{\sqrt{2}} [\hat k_\mathrm{a} \pm  \hat k_\mathrm{b} + \theta_\mathrm{a}(x) \pm \theta_\mathrm{b}(x)], \nonumber \\
    \phi_{\pm}(x) &=         \frac{1}{\sqrt{2}} [\phi_\mathrm{a}(x) \pm \phi_\mathrm{b}(x)],  \label{Eqn:theta_pm_phi_pm}
\end{align}
where the operators $\hat k_\gamma$ are absorbed by the fields $\hat \theta_{\pm}$, the bosonization of \eq{Eqn:Zoller_model} can be expressed as

\begin{align}
    H_0 &\sim  \frac{v_\mathrm{F}}{2 \pi} \sum_{s = {\pm}} \int_{0}^{\tilde{L}} :\left \{ [\partial_x \hat \theta_s(x)]^2 + [\partial_x \phi_s(x)]^2 \right\} : \mathrm{d} x, \label{Eqn:H_0_bosonized} \\
    H_W &\sim  \frac{4[\cos(2 k_\mathrm{F} a_0) - 1] W a_0}{(2 \pi \alpha)^2} \int_0^{\tilde{L}} \cos(\sqrt{8} \hat \theta_-) \mathrm{d} x. \label{Eqn:H_W_bosonized}
\end{align}
The Fermi velocity $v_\mathrm{F} = 2 t a_0 \sin(k_\mathrm{F} a_0)$ and the Fermi vector $k_\mathrm{F} = \frac{\nu  \pi}{a_0}$ depend on the filling fraction $\nu$.

Since the $+$ and $-$ fields commute, the Hamiltonian $H_0 + H_W$ decouples into a Sine-Gordon model and a free gapless bosonic theory. The free theory in the symmetric sector leads to a closing of the excitation gap $\propto 1 / \tilde{L}$. For sufficiently large values of $W$, spontaneous breaking of the wire parity symmetry $P_\mathrm{a} = (-1)^{N_\mathrm{a}}$ occurs in the antisymmetric sector, pinning the value of $\hat \theta_-$ to one of the two minima of the cosine \cite{Tu_et_al, Fidkowski_et_al, Halperin_Sarma}. The location of these minima depends on the sign of $W$, the two degenerate GS are thus characterized by $\hat \theta_- \approx \pm \pi / \sqrt{8}$ for $W<0$ and $\hat \theta_- \approx 0, \pi / \sqrt{2}$ for $W > 0$. More accurately, one should think of the two \gls{gs} being distinguished by the order parameter $\sin(\sqrt{2} \hat \theta_-)$ for $W < 0$ and $\cos(\sqrt{2} \hat \theta_-)$ for $W > 0$.

The parity operator $P_\mathrm{a}$ anticommutes with $e^{i \hat k_\mathrm{a}}$ and thus roughly speaking shifts $\hat k_\mathrm{a}$ by $\pi$, which corresponds to a shift of $\hat \theta_-$ by $\pi / \sqrt{2}$. Therefore, $P_\mathrm{a}$ exchanges the the two symmetry broken \gls{gs} characterized by $\hat \theta_- \approx \theta_1, \theta_2$, rendering their symmetric and anti-symmetric superposition parity eigenstates \cite{Fidkowski_et_al}:
\begin{align}
    \ket{P_\mathrm{a} = \pm 1} = \frac{1}{\sqrt{2}} \left [\ket{\hat \theta_- \approx \theta_1} \pm \ket{\hat \theta_- \approx \theta_2} \right]. \label{Eqn:parity_states}
\end{align}
These states correspond to the Majorana zero modes. Note that the above analysis is valid for \gls{obc}, in the case of \gls{pbc} only one of the two parity eigenstates is compatible with the boundary conditions on the fields $\phi_-$, $\hat \theta_-$ \cite{Fidkowski_et_al}.

\subsection{Bosonization of flux tunneling}
To investigate how and under which circumstances the symmetry-breaking term $H_\phi$ may split the GS degeneracy, we start by expressing it through the chiral fermionic fields in the continuum limit
\begin{align}
    H_\phi &\sim  r \int_{0}^{\tilde{L}} \left\{ e^{2 \pi i \phi x/a_0} \left[ \psi_\mathrm{R, a}^\dagger \psi_\mathrm{R, b} + \psi_\mathrm{L, a}^\dagger\psi_\mathrm{L, b} \right .  \right . \nonumber                                                      \\
                & \quad\left. \left.  +\, e^{-2i k_\mathrm{F} x} \psi_\mathrm{R, a}^\dagger \psi_\mathrm{L, b}+  e^{2 i k_\mathrm{F} x} \psi_\mathrm{L, a}^\dagger \psi_\mathrm{R, b}\right] + \text{H.c.} \right \}\mathrm d x, \label{Eqn:H_phi_continuum}
\end{align}
where we have suppressed the $x$-dependence of the fields for brevity. The necessity for commensurate flux can already be read off the above expression: rapidly oscillating terms will integrate out of the effective low-energy theory \cite{Giamarchi}, therefore we are left with interchain \gls{fs} for $\phi = 0$. At commensurate flux $\phi = \pm \nu$, the prefactor of one of the two interchain \gls{bs} terms becomes constant instead since  $k_\mathrm{F} = \frac{\nu  \pi}{a_0}$, making it the only relevant contribution to \eq{Eqn:H_phi_continuum}. In the following, we derive from the bosonization picture that only the \gls{bs} term will open a bulk gap and is useful for state preparation.

\subsubsection{Regular tunneling at $\phi = 0$}
The application of \eq{Eqn:Bosonization_identity} yields for the case without flux
\begin{align}
    H_{\phi = 0} &\sim  \frac{r(-1)^{N_\mathrm{a}}}{2 \pi \alpha} \int_{0}^{\tilde{L}}  \left[e^{i \frac{\pi}{\tilde{L}} (N_\mathrm{b}  - N_\mathrm{a} + 1) x} e^{i\sqrt{2}  [\hat \theta_-(x) + \phi_-(x)]} \right . \nonumber \\
                      & \quad+ \left . e^{- i \frac{\pi}{\tilde{L}} (N_\mathrm{b}  -  N_\mathrm{a} + 1) x} e^{i\sqrt{2} [\hat \theta_-(x) - \phi_-(x)]} - \text{H.c.} \right] \mathrm d x. \label{Eqn:H_phi_0_bosonized}
\end{align}
We retain the finite-size terms $\propto \frac{1}{\tilde{L}}$ because they will provide insight into the edge physics. As a consequence of
$\theta_-$ being pinned to a fixed value, the field $\phi_-$ is totally disordered in the bulk of the GS. The GS expectation value of the associated Hamiltonian density $\mathcal H_{\phi = 0}(x)$ is thus zero far away from the ends, therefore it cannot split the degeneracy \cite{Tu_et_al, Halperin_Sarma}. Close to the ends however, OBC enforce $\phi_- = 0$ \cite{Bosonization_OBC} and we may write

\begin{align}
    \mathcal H_{\phi = 0}(x \approx 0) &= r \frac{2 i}{\pi \alpha} (-1)^{N_\mathrm{a}} \sin[\sqrt{2} \hat \theta_-(x)], \nonumber \quad \\
    \mathcal H_{\phi = 0}(x \approx \tilde{L}´) &= -r \frac{2 i}{\pi \alpha} (-1)^{N_\mathrm{b}} \sin[\sqrt{2} \hat \theta_-(x)],
\end{align}
where the finite size term generates a relative minus sign between the contributions of the two ends if the total particle number $N_\mathrm{tot} = N_\mathrm{a} + N_\mathrm{b}$ is even. In conclusion, single-particle tunneling near the ends can only split the GS degeneracy for $W < 0$, but even then the contributions from both ends will cancel for even $N_\mathrm{tot}$. We give an argument that this cancellation happens on an exact level based on inversion symmetry in Appendix \ref{App:Sec:Bosonization}. Finally, we stress that while $H_{\phi = 0}$ may split the exponential degeneracy between the ground states in some cases, it cannot lift the finite size gap as it does not couple to the symmetric sector of the bosonized theory, cf. \eq{Eqn:H_0_bosonized}.

\begin{figure}[htp!]	
{
        \vbox to 0pt {
                \raggedright
                \textcolor{white}{
                    \subfloatlabel[1][Fig:DMRG_data_bosonization_phi_0_a]
                    \subfloatlabel[2][Fig:DMRG_data_bosonization_phi_0_b]
                }
            }
    } 
    \includegraphics{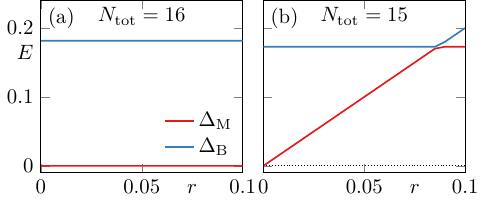}
\caption{The gaps $\Delta_\mathrm{M} = E_1 - E_0$ and $\Delta_\mathrm{B} = E_2 - E_0$ as a function of $r$ for a system size of $L = 24$ with \gls{obc}. Other parameters are $t = 1$, $W = -1.8$, $\mu = 0$, and $\phi = 0$. (a) Result for $N_\mathrm{tot} = 16$ particles, corresponding to exactly $\nu = 1/3$ and even total parity. (b) Result for $N_\mathrm{tot} = 15$ particles, corresponding to $\nu \approx 1/3$ and odd total parity.}\label{Fig:DMRG_data_bosonization_phi_0}
\end{figure}

The subtle effects of the total fermion parity and the sign of $W$ predicted by our constructive bosonization approach can be readily observed in DMRG simulations as we demonstrate in Fig.~\ref{Fig:DMRG_data_bosonization_phi_0} for a system of $L = 24$ sites with parameters $W = -1.8$, $t = 1$, $\mu = 0$, and $\phi = 0$. Fig.~\ref{Fig:DMRG_data_bosonization_phi_0_a} shows the gaps $\Delta_\mathrm{M}$ and $\Delta_\mathrm{B}$ as a function of the interchain hopping amplitude $r$ for even particle number $N_\mathrm{tot} = 16$. Neither the degeneracy of the Majorana modes nor the bulk gap are affected by moderate values of $r$. This is contrasted by the data in Fig.~\ref{Fig:DMRG_data_bosonization_phi_0_b}, where odd total fermion parity at a similar filling fraction $\nu \approx 1/3$ is achieved by taking $N_\mathrm{tot} = 15$ particles. Then, the \gls{gs} degeneracy is immediately split by interchain hopping, but without lifting the finite-size gap to the bulk states. As predicted by the bosonization analysis, we observe no \gls{gs} splitting for either odd or even total parity if $W > 0$ (not shown in Fig.~\ref{Fig:DMRG_data_bosonization_phi_0}). 

\subsubsection{Tunneling with commensurate flux at $\phi = \nu$}
At commensurate flux $\phi = \nu$, \eq{Eqn:H_phi_continuum} effectively reduces to interchain \gls{bs}, which bosonizes to
\begin{align}
    H_{\phi = \nu} &\sim  \frac{-i r (-1)^{N_\mathrm{a}}}{\pi \alpha}  \int_{0}^{\tilde{L}} \left \{ \cos \left[ \sqrt{2} \hat \theta_-(x) \right] \right . \nonumber                                 \\
                        & \quad \times  \sin \left [ \frac{\pi}{\tilde{L}} (\Delta N_\mathrm{tot} + 1 - 2 \nu) x + \sqrt{2} \phi_+(x) \right]   \nonumber                                                   \\
                        & \quad + \sin \left[ \sqrt{2} \hat \theta_-(x) \right]   \nonumber                                                                                                         \\
                        & \quad \times \left.  \cos \left [ \frac{\pi}{\tilde{L}} (\Delta N_\mathrm{tot} + 1 - 2 \nu) x +  \sqrt{2} \phi_+(x) \right] \right \} \mathrm d x, \label{Eqn:H_phi_nu_bosonized}
\end{align}
where $\Delta N_\mathrm{tot} = N_\mathrm{a} + N_\mathrm{b} - 2 \nu L$ is the deviation from filling fraction exactly $\nu$ of the underlying lattice model. Depending on the sign of $W$, either the term in the first or second line will distinguish the two symmetry-broken \gls{gs}, which is why a splitting of the degeneracy is expected in either case. Furthermore, the finite-size terms will generally not conspire to an integer multiple of $\pi$ at $x \approx \tilde{L}$, meaning that there will be no exact cancellation of the contributions to \gls{gs} splitting from the two ends in contrast to the case $\phi = 0$, even though the result is still sensitive to the total fermion parity. To explain the effect of \eq{Eqn:H_phi_nu_bosonized} on the bulk properties of the system, the finite-size terms can be neglected and the $\hat \theta_-$ term can be replaced by a mean-field value due to the large gap in the antisymmetric sector (see Appendix \ref{App:Sec:Bosonization} for more details). Then, a Sine-Gordon theory remains in the symmetric sector, which subsequently flows to a massive phase under RG \cite{Giamarchi} and implies the formation of a finite gap in the bulk.

\begin{figure}[htp!]	 
{
        \vbox to 0pt {
                \raggedright
                \textcolor{white}{
                    \subfloatlabel[1][Fig:DMRG_data_bosonization_phi_nu_a]
                    \subfloatlabel[2][Fig:DMRG_data_bosonization_phi_nu_b]
                }
            }
    } 
    \includegraphics{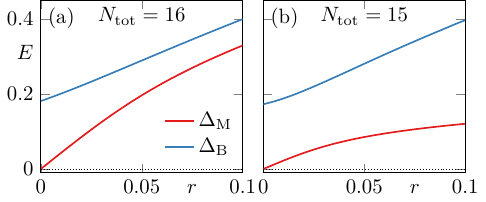}
\caption{The gaps $\Delta_\mathrm{M} = E_1 - E_0$ and $\Delta_\mathrm{B} = E_2 - E_0$ as a function of $r$ for a system size of $L = 24$ with \gls{obc}. Other parameters are $t = 1$, $W = -1.8$, $\mu = 0$, and $\phi = 1/3$. (a) Result for $N_\mathrm{tot} = 16$ particles, corresponding to exactly $\nu = 1/3$ and even total parity. (b) Result for $N_\mathrm{tot} = 15$ particles, corresponding to $\nu \approx 1/3$ and odd total parity.}\label{Fig:DMRG_data_bosonization_phi_nu}
\end{figure}

Again, we compare the field-theoretical prediction with \gls{dmrg} data in Fig.~\ref{Fig:DMRG_data_bosonization_phi_nu} for a system of size  $L = 24$ with \gls{obc} and $t = 1$, $W = -1.8$, $\mu = 0$, and $\phi = \nu = 1/3$. Fig.~\ref{Fig:DMRG_data_bosonization_phi_nu_a} shows the gaps $\Delta_\mathrm{M}$ and $\Delta_\mathrm{B}$ as a function of $r$ for the even parity case $N_\mathrm{tot} = 16$ and $\Delta N_\mathrm{tot} = 0$, clearly indicating the splitting of the \gls{gs} degeneracy and the formation of a bulk gap. Comparing this to the case of odd parity at similar filling fraction $N_\mathrm{tot} = 15$, $\Delta N_\mathrm{tot} = 1$, and $\nu \approx 1/3$ in Fig.~\ref{Fig:DMRG_data_bosonization_phi_nu_b}  shows a similar behavior of the bulk gap but a smaller \gls{gs} splitting. These numerical results precisely reflect \eq{Eqn:H_phi_nu_bosonized}: the bulk gap should not depend on slowly varying finite-size terms, the \gls{gs} splitting however is expected to be carried by the two ends of the chain \cite{Tu_et_al, Halperin_Sarma} and is thus sensitive to the relative sign between their respective contributions. For negative $W$, only the second term of \eq{Eqn:H_phi_nu_bosonized} contributes to the splitting; after taking into account the boundary conditions on the field in the symmeric sector $\phi_+(0) = \phi_+(\tilde{L}) = 0$, we find that the right end is weighted by  a factor $\cos[\pi(\Delta N_\mathrm{tot} + 1/3)]$ relative to the left end, amounting to an amplification or cancellation for $\Delta N_\mathrm{tot}$ even and odd, respectively. 

We close this section by emphasizing that while $H_\phi$ breaks the protecting $\mathbb{Z}_2$ symmetry for any value of $\phi$, the simple case of $H_{\phi = 0}$ only splits the topological \gls{gs} degeneracy under certain circumstances and never takes the system away from the critical point. Thus, tunneling with commensurate flux $H_{\phi = \nu}$ provides the necessary ``mass'' term for the efficient preparation of the critical state \cite{crit_state_prep_1, crit_state_prep_2, crit_state_prep_3, Review_QPT_Dziarmaga, crit_state_prep_4}. 
To further exemplify this fundamental difference between the cases of $\phi = 0$ and $\phi = \nu$, we calculate the superconducting correlator $c_{\mathrm{a}, 1} c_{\mathrm{a}, 2} c_{\mathrm{a}, j}^\dagger c_{\mathrm{a}, j+1}^\dagger$ for both in Fig.~\ref{Fig:cooper_correlator_a}. There, we set $r = 0.05$ and choose a system size of $L = 96$ at filling exactly $\nu = 1/3$. At zero flux, the algebraic \gls{sc} order clearly persists, hallmarking the criticality of the state. By contrast, the commensurate flux $\phi = \nu$ leads to the expected breakdown of criticality, resulting in an exponential decay of the \gls{sc} correlator. This is complemented by data on the associated Schmidt spectrum arising from a spatial bipartition plotted as a function of $r$ in Fig.~\ref{Fig:cooper_correlator_b}. Here, we resort to \gls{pbc} to avoid complications resulting from \gls{gs} degeneracy. The two-fold degeneracy expected from a $\mathbb Z_2$ protected topological phase in \gls{1d} \cite{1d_topological_phases_1, 1d_topological_phases_2, 1d_topological_phases_3} is clearly split for $\phi = \nu$ while it remains intact at $\phi = 0$, even though both cases break the protecting symmetry. This behavior is consistent with our analysis of the \gls{gs} degeneracy splitting for the case of even parity presented earlier, since the \gls{gs} degeneracy for \gls{obc} is closely related to the degeneracy of the entanglement spectrum across a spatial cut.

\begin{figure}[htp!]	 
{
        \vbox to 0pt {
                \raggedright
                \textcolor{white}{
                    \subfloatlabel[1][Fig:cooper_correlator_a]
                    \subfloatlabel[2][Fig:cooper_correlator_b]
                }
            }
    } 
    {\includegraphics[trim={0.75cm 0cm 0.75cm 0.cm}, width=0.85\linewidth]{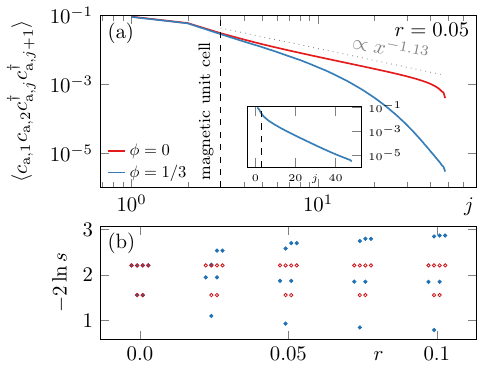}}
    \caption{In (a), the Cooper pair correlator  $c_{\mathrm{a}, 1} c_{\mathrm{a}, 2} c_{\mathrm{a}, j}^\dagger c_{\mathrm{a}, j+1}^\dagger$ on the a-wire with and without flux $\phi$ close to the end of the protocol for $r=0.1$ and a system size of $L = 96$ with \gls{obc} and $N_\mathrm{tot} = 64$ particles corresponding to $\nu = 1/3$ is shown. Other parameters are $t = 1$, $W = -1.8$, and $\mu = 0$. While the correlation function decays algebraically without flux, this order is destroyed with flux. This is complemented by the Schmidt spectrum for a state with \gls{pbc} in (b) indicating a breaking of the topological phase as the Schmidt weights split.}\label{Fig:cooper_correlator}
\end{figure}

\section{Conclusion}\label{Sec:conclusion}
We have investigated the preparation of a critical Majorana phase in a two-leg ladder with total particle number conservation, where the wire parity $P_\mathrm{a} = (-1)^{N_\mathrm{a}}$ acts as the protecting $\mathbb Z_2$ symmetry. The key ingredient of our protocol is a symmetry-breaking perturbation in the form of an inter-wire tunneling that threads an artificial magnetic flux commensurate with the filling fraction through each plaquette of the ladder, thereby acting as a ``mass'' term that immediately gaps out the critical phase. This additional term is feasible with existing experimental techniques. We argue how our protocol achieves an asymptotically optimal scaling of preparation time linear with system size, and provide \gls{mpste} data to substantiate our findings. For comparison, we also study a symmetry-respecting preparation approach that is found to require quadratic ramp times.

The theoretical backbone of this work is a constructive bosonization analysis of the Majorana Hamiltonian and the flux hopping term, taking into account particle-number dependent finite-size terms and carefully constructed Klein factors. On this basis, we demonstrate why previously discussed symmetry-breaking terms, e.g., a regular inter-wire tunneling without flux, are not helpful for state preparation, by contrast to the commensurate flux term. Moreover, the bosonized theory is able to resolve subtle effects such as qualitative differences between systems with odd and even total fermion parity. The field-theoretical predictions are fully confirmed through \gls{dmrg} simulations.

Finally, we would like to discuss the relation to some previous work on topological state preparation. First, we note that the dissipative preparation as the steady state of a quantum master equation of a \gls{1d} Majorana phase in a number-conserving setting has also been discussed in Ref.~\cite{Dissipative_Majorana_Preparation}. There, the typical preparation time is expected to scale with system size as $\tau_\mathrm{tot}\propto L^2$. Second, a detailed proposal for the preparation of an integer Chern insulator state through the augmentation with the inverted topological phase has been put forward in Ref.~\cite{Barbarino_et_al}. As topological superconductors at mean field level, including the Kitaev chain, formally fall into the class of invertible topological phases, one may consider extending these ideas to the state preparation of Majorana phases.  However, we note that our present system operates at a critical point in a strongly interacting regime, which at least challenges any mean field picture of topological superconductivity, such that the results of Ref.~\cite{Barbarino_et_al} cannot be adapted directly. 

\acknowledgments
We would like to thank Sebastian Diehl for discussions. We acknowledge financial support from the German Research Foundation (DFG) through the Collaborative Research Centre (SFB 1143, project ID 247310070) and the Cluster of Excellence ct.qmat (EXC 2147, project ID
390858490). Our numerical calculations were performed on resources at the TU Dresden Center for Information Services and High Performance Computing (ZIH).

\vspace{15pt}
\section*{DATA AVAILABILITY}
The raw data for each data figure including a script to generate it is available on Zenodo \cite{Zenodo}.

\appendix

\onecolumngrid

\section{More details on critical state preparation}\label{App:Sec:crit_state_prep}

\subsection{More details on the asymptotically linear scaling}

To argue for the linear scaling of our protocol, we start from an estimate for the adiabatic preparation time $\tau_\mathrm{tot}$ of a critical system with dynamical and critical exponents $z_\mathrm{c} = \nu_\mathrm{c} = 1$
\begin{align}
\tau_\mathrm{tot} \gg L^{1 + \frac{1}{p}} \label{EqnApp:Prep_time_scaling}
\end{align}
assuming that the critical point is approached as a power-law 
\begin{align}
r_p(\tau) = r_{0} |1 -  \tau / \tau_\mathrm{tot}|^p \label{EqnApp:power_law_ramp}
\end{align}
with $\tau \in [0,\tau_\mathrm{tot}]$, where $r_p$ is the mass-term that controls the distance to the critical point in the sense of opening a finite gap above the \gls{gs} of the otherwise critical system. This estimate has been derived  in the context of the Kibble-Zurek mechanism from the scaling behavior of the correlation length close to the critical point \cite{crit_state_prep_3, Review_QPT_Dziarmaga} under the assumption that the adiabatic ramp starts with the exact \gls{gs} of the Hamiltonian at $r_p(\tau_\mathrm{tot} = 0) = r_{0}$. However, since the rate of change  
\begin{align}
|\dot{r}_p(\tau = 0)| = r_{0} p / \tau_\mathrm{tot}  \label{EqnApp:roc_t_0}
\end{align}
at this point is not bounded, naively taking the limit $p \to \infty$ to obtain a protocol with linear scaling of preparation time will lead to non-adiabatic processes at the beginning of the ramp. Hence, given a value of $p$, \eq{EqnApp:Prep_time_scaling} is only valid for large enough system sizes where $p / \tau_\mathrm{tot}(L)$ is small enough. Assuming that \eq{EqnApp:roc_t_0} and \eq{EqnApp:Prep_time_scaling} pose the main constraints for the adiabaticity of the ramp, a slow enough sub-linear increase of the power law $\lim_{L \to \infty}p(L) = \infty$ should: $(i)$ ensure \eq{EqnApp:roc_t_0} is no concern because $\lim_{L\to \infty} p(L) / \tau_\mathrm{tot}(L) = 0$ and $(ii)$ enable asymptotically linear scaling $\tau_\mathrm{tot}(L) \propto L^{1 + \frac{1}{p(L)}} \approx L$. 

The data in Fig.~\ref{Fig:fidelity_data} shows \gls{mpste} data for system sizes $L = 24, 48, 72, 96$ and powers $p = 2.2, 2.7, 3, 3.2$, respectively, that we determined numerically to yield optimal fidelity for given preparation time $\tau_\mathrm{tot}$. The curves do not fully collapse under linear scaling, which is expected due to subleading corrections $\mathcal O (1/L)$ to the asymptotic scaling $\tau_\mathrm{tot}(L) \propto L$ and also numerical limitations of the \gls{mpste} method, which increase with system size and duration of the simulated time evolution. Still, the data clearly illustrates the main point that $p$ can be increased with system size, which implies asymptotically linear scaling by \eq{EqnApp:Prep_time_scaling}.

\subsection{Single-particle correlations in the prepared states and preparation of a parity eigenstate}
\begin{figure}[htp!]	 
\includegraphics[width=0.9\textwidth]{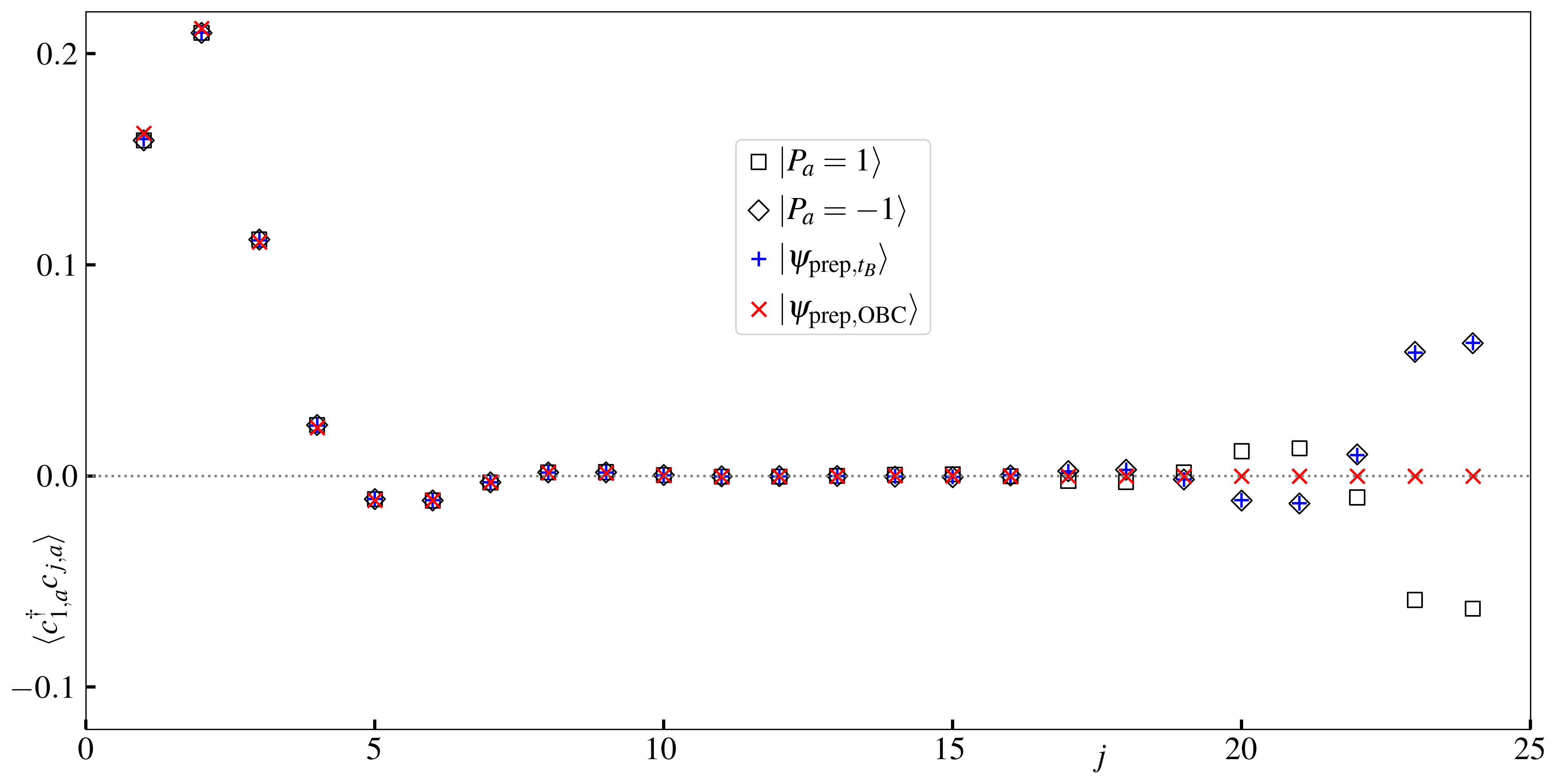}
\caption{Single particle correlation function $\langle c^\dagger_{1,a} c_{j,a} \rangle$ for various \gls{gs} of the Hamiltonian \eq{Eqn:Zoller_model} with parameters $t = 1$, $W = -1.8$, system size $L = 24$ and \gls{obc}. The square and diamond represent the correlations in eigenstates of $P_\mathrm{a} = (-1)^{N_\mathrm{a}}$ corresponding to the Majorana zero modes. The red crosses are the correlations in the state $\ket{\psi_{\mathrm{prep}, \text{\gls{obc}}}}$ that was prepared using the flux-assisted protocol with \gls{obc} as described by \eq{Eqn:time_dependence_p1}. Finally, the correlations in $\ket{\psi_{\mathrm{prep}, t_B}}$ that was prepared with an additional boundary hopping $t_B$ (cf.~\eq{EqnApp:H_tau}) are shown as blue pluses. \label{Fig:Correlators}}
\end{figure}
For any ground state of the Majorana Hamiltonian \eq{Eqn:Zoller_model}, the single-particle correlator $G_{a/b}(j,l) = \bra{\gls{gs}}c^\dagger_{j, a/b} c_{l, a/b} \ket{\gls{gs}}$ on either wire exhibits a superconducting gap in the bulk, i.e. exponential decay with $|j-l|$ for positions $j,l$ chosen far away from the boundary. If one of the \gls{gs}s $\ket{P_\mathrm{a} = \pm 1}$ [cf. \eq{Eqn:parity_states}] that are simultaneously eigenstates of the subwire parity operator $P_\mathrm{a} = (-1)^{N_\mathrm{a}}$ are chosen, non-local correlations between the edges can be observed, see Fig.~\ref{Fig:Correlators}. These special ground states are the Majorana zero modes of the system.

Following the state preparation protocol with \gls{obc} as we do in the main text yields a ground state that is roughly an eigentstate of the $\hat \theta_-$ operator and thus an evenly weighted superposition of the two parity eigenstates [cf. again \eq{Eqn:parity_states}]. Since the correlations with the opposite edge have opposite sign for opposite parities, they will generally cancel out in the prepared state. To prepare a parity eigenstate with the flux-assisted protocol, one can start from the ground state of the Hamiltonian with commensurate flux hopping $r = 0.1$ and \gls{pbc} on the chain hopping $t$ and then ramp down the boundary terms together with the flux hopping. This corresponds to the time-dependent Hamiltonian

\begin{align}
H_0 + H_W + r(\tau) \sum_{j = 1}^L  \left[ e^{2 \pi i \phi j}c_{\mathrm{a},j}^\dagger c_{\mathrm{b},j} + e^{-2 \pi i \phi j}c_{\mathrm{b},j}^\dagger c_{\mathrm{a},j} \right] - t_{B}(\tau) [c_{L,a}^\dagger c_{1,a} + c_{L,b}^\dagger c_{1,b} + \text{H.c.}] , \label{EqnApp:H_tau} 
\end{align}
with $H_0$ and $H_W$ as per \eq{Eqn:Zoller_model} and
\begin{align}
r_p(\tau) = r_{0} |1 -  \tau / \tau_\mathrm{tot}|^p, \quad t_{B}(\tau) = t |1 -  \tau / \tau_\mathrm{tot}|, \label{EqnApp:power_law_ramp_t_B}
\end{align}
where $r_0 = 0.1$. Generally, the boundary hopping $t_{B}(\tau)$ will energetically favor one of the two parity eigenstates and choosing a linear ramp should make the energetic penalty dominant enough to end up with a parity eigenstate without generating excited states and compromising the overall fidelity. 

We simulated the above protocol for $\tau_\mathrm{tot} = 50$ and $p = 2.2$, which yields a state $\ket{\psi_{\mathrm{prep}, t_B}}$ with a total fidelity $F_{t_B} = |\braket{P_\mathrm{a} = 1|\psi_{\mathrm{prep}, t_B}}|^2 + |\braket{P_\mathrm{a} = -1|\psi_{\mathrm{prep}, t_B}}|^2 \approx 0.998$ and an overlap $|\braket{P_\mathrm{a} = -1|\psi_{\mathrm{prep}, t_B}}|^2 \approx 0.987$ with an individual parity eigenstate. For comparison, the protocol with purely \gls{obc} as simulated in the main text (cf.~Fig.~\ref{Fig:fidelity_data}) results in a state  $\ket{\psi_{\mathrm{prep}, \text{\gls{obc}}}}$ with a total fidelity of $F_\text{{\gls{obc}}}\approx 0.9991$ but an evenly distributed overlap of $\approx 0.5$ with $\ket{P_\mathrm{a} = 1}$ and $\ket{P_\mathrm{a} = -1}$ when choosing $\tau_\mathrm{tot} = 50$ and $p = 2.2$. 
Adding the boundary term introduces additional entanglement that prevents us from simulating larger system sizes, but the result for $L = 24$ indicates that the additional boundary term will select a parity eigenstate without substantially diminishing the fidelity of the prepared state. We show the resulting correlation function for $\ket{\psi_{\mathrm{prep}, t_B}}$ and $\ket{\psi_{\mathrm{prep}, \text{\gls{obc}}}}$ in Fig.~\ref{Fig:Correlators}, which illustrates that $\ket{\psi_{\mathrm{prep}, t_B}}$ exhibits single-particle edge correlations similar to $\ket{P_\mathrm{a} = -1}$ while they are absent for $\ket{\psi_{\mathrm{prep}, \text{\gls{obc}}}}$ as expected.

As a final remark, we would like point out that if the setup is intended to be used for quantum information processing, some means of measuring and braiding the Majorana modes has to be devised anyway. Then, one can prepare any \gls{gs} of the Hamiltonian and make it collapse into either of the Majorana modes by measuring the subsytem parity $P_\mathrm{a}$. While experimentally challenging, this is an order $\mathcal O(L)$ effort in principle and does not change the $\mathcal O(L)$ scaling of the protocol.

\section{Stability against disorder} \label{App:Sec:stability_perturbations}

\begin{figure}[htp!]	 
{
        \vbox to 0pt {
                \raggedright
                \textcolor{white}{
                    \subfloatlabel[1][Fig:disorder_data_appendix_a]
                    \subfloatlabel[2][Fig:disorder_data_appendix_b]
                }
            }
    }
    \includegraphics[width=0.9\textwidth]{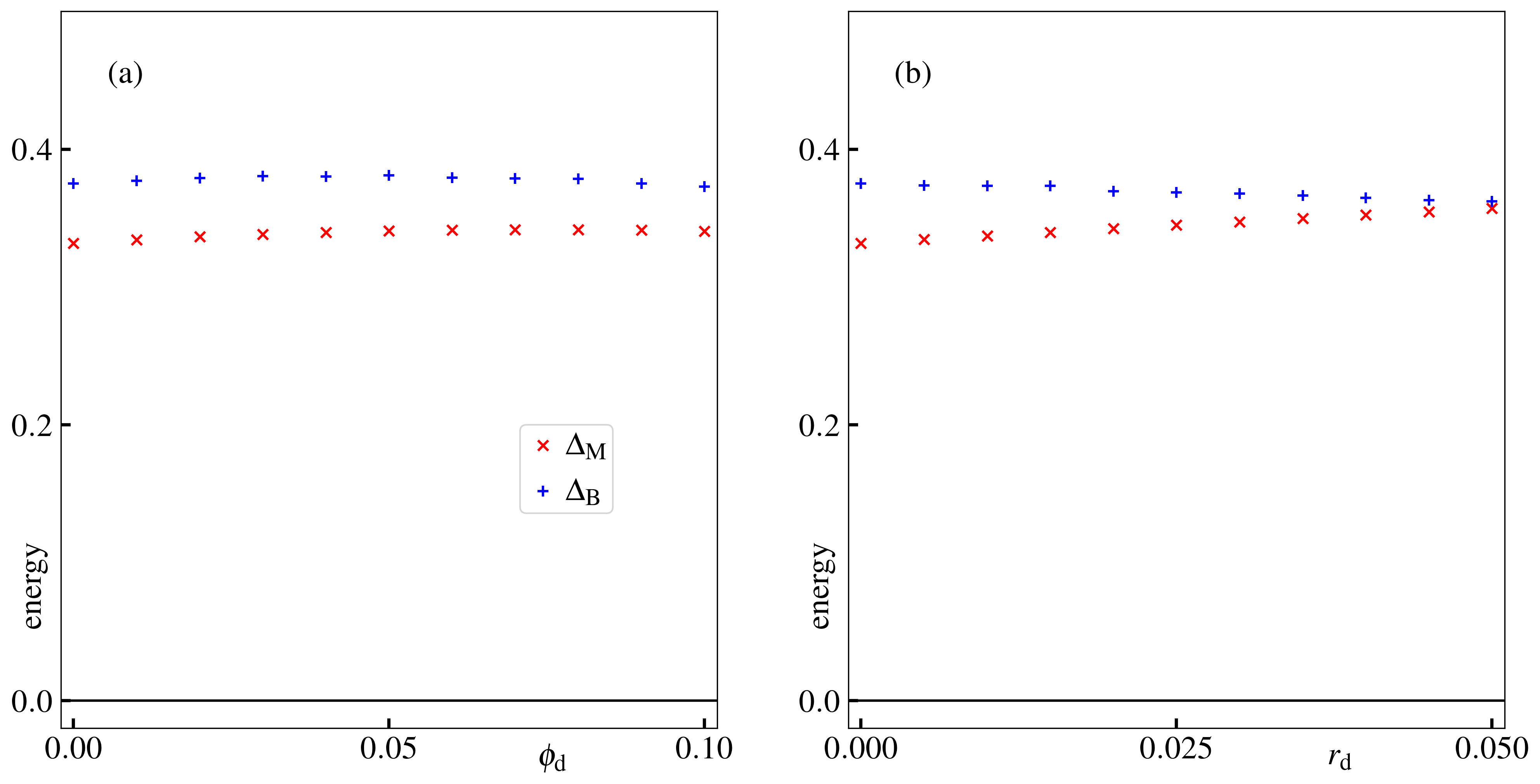}
\caption{The gaps $\Delta_\mathrm{M} = E_1 - E_0$ and $\Delta_\mathrm{B} = E_2 - E_0$ for a system of $L = 48$ and \gls{obc} populated by $N_\mathrm{tot} = 32$ particles corresponding to $\nu = 1/3$ with parameters $t = 1$, $W = -1.8$, $\mu = 0$, $r = 0.1$, and $\phi = 1/3$ in the presence of disorder on the transverse hopping as introduced in \eq{EqnApp:H_phi_disorder}. (a) Gaps for $r_\mathrm{d}= 0$ and increasing $\phi_\mathrm{d}$. (b) Gaps for $\phi_\mathrm{d}= 0$ and increasing $r_\mathrm{d}$.}
\end{figure}

Here we show that the preparation scheme is stable against small imperfections in the realization of exactly commensurate flux and homogeneous hopping amplitude. We consider adding a small random variation to either the amplitude or the phase in the transverse hopping term:
	
\begin{align}
H_\phi^{\text{dis}} &= \sum_{j = 1}^L  (r + \Delta_{r, j})  \left[ e^{2 \pi i [\phi j + \Delta_{\phi, j}]}c_{\mathrm{a},j}^\dagger c_{\mathrm{b},j} + e^{-2 \pi i [\phi j + \Delta_{\phi, j}]}c_{\mathrm{b},j}^\dagger c_{\mathrm{a},j} \right], \label{EqnApp:H_phi_disorder} 
\end{align}
where $ \Delta_{r, j}$ and $\Delta_{\phi, j}$ are randomly and independently drawn from the uniform distribution on the interval $[-r_\mathrm{d}, r_\mathrm{d}]$, $[-\phi_\mathrm{d}, \phi_\mathrm{d}]$, respectively. Fig.~\ref{Fig:disorder_data_appendix_a} shows the resulting energy gaps for random fluctuations up to $\phi_\mathrm{d} = 0.1$ around the commensurate value of $\phi = 1/3$ while $r = 0.1$ is fixed. Clearly, both the splitting of the \gls{gs} degeneracy as well as the formation of the bulk gap persist. In Fig.~\ref{Fig:disorder_data_appendix_b}, we investigate random fluctuations up to $r_\mathrm{d} = 0.05$ around the value $r = 0.1$ at fixed $\phi = 1/3$, for which the \gls{gs} degeneracy as well as the bulk gap are also stable. 

The observed stability against random fluctuations is expected, since quickly fluctuating terms should integrate out of the low-energy effective theory \cite{Giamarchi}, as we also discuss in Appendix \ref{App:Sec:Bosonization}. Because the formation of a bulk gap is the only relevant function of the flux hopping for the presented protocol, we conclude that small experimental imperfections should not impede the feasibility of our strategy.

\section{Bosonization} \label{App:Sec:Bosonization}
\subsection{Fermionic continuum fields for \gls{obc}}

We consider fermionic fields defined on the interval $[0, \tilde{L}]$ with \gls{obc} following Ref.~\cite{Bosonization_OBC}, which obey the boundary conditions
\begin{align}
\psi_\gamma(0) = \psi_\gamma(\tilde{L}) = 0, \quad \gamma = \mathrm{a, b}. \label{EqnApp:def_OBC}    
\end{align}
The fields can be expanded in Fourier modes
\begin{align}
\psi_\gamma(x) = \sqrt{\frac{2}{\tilde{L}}} \sum_{n = 1}^\infty \sin(k_n x) c_{\gamma, n} \label{EqnApp:Mode_expansion_OBC}
\end{align}
where $c_n$ annihilates a particle with momentum $k_n = n \pi / \tilde{L}$. Note that there are only positive momenta and accordingly only a single Fermi point at some $k_\mathrm{F} > 0$. Now, slowly varying chiral fields can be defined 
\begin{align}
\psi_\mathrm{\gamma, R}(x) = -\frac{i}{\sqrt{2 \tilde{L}}} \sum_{n = 1}^\infty e^{i (k_n - k_F) x}c_{\gamma,n}, \quad \psi_\mathrm{\gamma, L}(x) = \frac{i}{\sqrt{2 \tilde{L}}} \sum_{n = 1}^\infty e^{-i (k_n - k_F)  x}c_{\gamma, n}, \label{EqnApp:Mode_expansion_Psi_RL_OBC}    
\end{align}
Later on, the approximation of letting the sums run to $- \infty$ will be made. This is the usual approximation taken in bosonization schemes, justified by the assumption that all the relevant physics take place close to the Fermi surface where $n \approx \tilde{L} k_\mathrm{F} / \pi$. The L/R fields are composed of the same set of momentum operators and related by 
\begin{align}
\psi_\mathrm{\gamma, L}(x) = - \psi_\mathrm{\gamma, R}(-x). \label{EqnApp:relation_Psi_RL_OBC}    
\end{align}
The fermionic field can be written in terms of L/R fields as 
\begin{align}
\psi_\gamma(x) = e^{i k_\mathrm{F} x} \psi_\mathrm{\gamma, R}(x)  + e^{-i k_\mathrm{F} x} \psi_\mathrm{\gamma, L}(x) = e^{i k_\mathrm{F} x} \psi_\mathrm{\gamma, R}(x)  - e^{-i k_\mathrm{F} x} \psi_\mathrm{\gamma, R}(-x).  \label{EqnApp:LR_decomposition}    
\end{align}

\subsection{Bosonization identity and commutators}
The R fields have periodicity $\tilde{L}' = 2 \tilde{L}$ and are therefore bosonized by a constructive bosonization approach for periodic fermion fields following Schönhammer \cite{Bosonization_Schoenhammer}
\begin{align}
\psi_\mathrm{a, R}(x) &= \frac{1}{\sqrt{2 \pi \alpha}} e^{i \hat k_\mathrm{a}} e^{i \frac{2\pi}{\tilde{L}'} \Delta N_\mathrm{a} x} e^{i \vartheta_\mathrm{a}(x)} = \frac{1}{\sqrt{2 \pi \alpha}} e^{i \hat k_\mathrm{a}} e^{i \frac{\pi}{\tilde{L}} \Delta N_\mathrm{a} x} e^{i \vartheta_\mathrm{a}(x)} = \frac{1}{\sqrt{2 \pi \alpha}} e^{i \frac{\pi}{\tilde{L}} [\Delta N_\mathrm{a} +1]x} e^{i [\hat k_\mathrm{a} + \vartheta_\mathrm{a}(x)]}, \nonumber \\
\psi_\mathrm{b, R}(x) &= \frac{(-1)^{N_\mathrm{a}}}{\sqrt{2 \pi \alpha}} e^{i \hat k_\mathrm{b}} e^{i \frac{2\pi}{\tilde{L}'} \Delta  N_\mathrm{b}x} e^{i \vartheta_\mathrm{b}(x)} = \frac{(-1)^{N_\mathrm{a}}}{\sqrt{2 \pi \alpha}} e^{i \hat k_\mathrm{b}} e^{i \frac{\pi}{\tilde{L}} \Delta  N_\mathrm{b}x} e^{i \vartheta_\mathrm{b}(x)} = \frac{(-1)^{N_\mathrm{a}}}{\sqrt{2 \pi \alpha}} e^{i \frac{\pi}{\tilde{L}} [\Delta  N_\mathrm{b}+ 1]x} e^{i[\hat k_\mathrm{b} +  \vartheta_\mathrm{b}(x)]}, \label{EqnApp:Bosonization_ID_2LL}
\end{align}
for which we formally extend the summation in \eq{EqnApp:Mode_expansion_Psi_RL_OBC} to $- \infty$. In the above equation, $\alpha$ plays the role of a regularization parameter, $\Delta  N_\gamma = N_\gamma - \tilde{L} k_\mathrm{F} / \pi$ counts the number of $\gamma = \mathrm{a, b}$ particles relative to the Fermi surface, and the Hermitian operators $\hat k_\gamma$ are conjugate to the particle number in the sense that 
\begin{align}
[N_\gamma, e^{\pm i \hat k_{\gamma'}}] = \mp e^{\pm i \hat k_\gamma}\delta_{\gamma, \gamma'} \quad \Leftrightarrow \quad (N_\gamma \pm \delta_{\gamma, \gamma'}) e^{\pm i \hat k_\gamma'} = e^{\pm i \hat k_\gamma'}  N_\gamma, \label{EqnApp:N_k_relation_2}
\end{align}
while they commute among themselves and with the fields $\vartheta_\gamma$. The operator $e^{\pm i \hat k_\gamma}$ represents the particle-number changing property of $\psi_\mathrm{\gamma, R}(x)$ and anticommutes with the associated parity $(-1)^{N_\gamma}$. This ensures the anticommutation of different fermion species in \eq{EqnApp:Bosonization_ID_2LL} and provides an explicit construction of Klein factors. As a side remark, some parts of the literature claim the stronger relation $[N_\gamma, \hat k_{\gamma'}] = i \delta_{\gamma, \gamma'}$, however, corrections arise on the level of constructive bosonization that only permit the weaker statement (see again \cite{Bosonization_Schoenhammer}). All results presented here are derived using the correct commutator.

To construct the phase fields, bosonic operators are defined from the Fourier components of the electron density as 
\begin{align}
b_{\gamma, n} = \frac{-i}{\sqrt{|n|}} \sum_{m} c^\dagger_{\gamma, n} c_{\gamma, n+m}, \quad b_{\gamma, n}^\dagger = \frac{i}{\sqrt{|n|}} \sum_{m} c^\dagger_{\gamma, n +m} c_{\gamma, n}, \quad \text{for } n > 0. \label{EqnApp:def_b_n}
\end{align}
whose bosonic commutation relations 
\begin{align}
[b_{\gamma, n},b_{\gamma', n'}] = [b_{\gamma, n}^\dagger,b_{\gamma', n'}^\dagger] = 0, \quad [b_{\gamma, n},b_{\gamma', n'}^\dagger] = \delta_{\gamma, \gamma'} \delta_{n, n'} \label{EqnApp:b_com}
\end{align}
follow immediately from the properties of the fermionic operators $c_{\gamma, n}$. The fields from \eq{EqnApp:Bosonization_ID_2LL} are then 
\begin{align}
\vartheta_\gamma(x) = \sum_{n > 0} \frac{e^{- \alpha q_n / 2}}{\sqrt{n}} \left[e^{i q_n x} b_{\gamma, n} +  e^{- i q_n x}b_{\gamma, n}^\dagger \right ] \label{EqnApp:def_vartheta_ab}
\end{align}
and the commutator has been derived in \cite{Bosonization_Delft} 
\begin{align}
[\vartheta_\gamma(x), \vartheta_{\gamma'}(y)] =&\; \; \delta_{\gamma,\gamma'} i [2\arctan[(x-y)/\alpha] - \pi (x-y) /  \tilde{L}], \nonumber \\
\stackrel{\mathclap{\alpha \to 0}}{=} & \;\; \delta_{\gamma,\gamma'} i \pi [\text{sign}(x-y) -(x-y) /  \tilde{L}] \quad \text{for } x,y \in [- \tilde{L}, \tilde{L}]. \label{EqnApp:com_vartheta_ab}
\end{align}
It is convenient to define the fields
\begin{align}
\theta_\gamma(x) &= \frac{\vartheta_\gamma(x) + \vartheta_\gamma(-x)}{2} = \sum_{n > 0} \frac{e^{ - \alpha q_n / 2} \cos(q_n x)}{\sqrt{n}} \left [b_{\gamma, n} + b_{\gamma, n}^\dagger \right], \nonumber \\ 
\phi_\gamma(x) &= \frac{\vartheta_\gamma(x) - \vartheta_\gamma(-x)}{2} = \sum_{n > 0} \frac{e^{ - \alpha q_n / 2} i\sin(q_n x)}{\sqrt{n}} \left [b_{\gamma, n} - b_{\gamma, n}^\dagger \right]. \label{EqnApp:def_theta_phi}
\end{align}
which are used to express the bosonization identity in the main text. The above equation makes clear that these fields have a periodicity of $2 \tilde{L}$ as well as the properties $\theta_\gamma(-x) = \theta_\gamma(x)$, $\phi_\gamma(-x) = -\phi_\gamma(x)$, and crucially $\phi_\gamma(0) = \phi_\gamma(\tilde{L}) = 0$. Their commutators are readily derived from \eq{EqnApp:b_com} and \eq{EqnApp:com_vartheta_ab}:

\begin{align}
[\theta_{\gamma}(x), \theta_{\gamma'}(y)] = [\phi_{\gamma}(x), \phi_{\gamma'}(y)]  = 0
\end{align}
and 
\begin{align}
[\theta_{\gamma}(x), \phi_{\gamma'}(y)] &= \delta_{\gamma,\gamma'} \frac{i}{2} \big \{ 2\arctan[(x-y)/\alpha] - 2\arctan[(x+y)/\alpha] + 2 \pi y / \tilde{L} \big\} ] \quad \text{for } x,y \in [0, \tilde{L}]. 
\end{align}
Assuming that $x$ and $y$ differ by a sufficiently large amount, we can approximate $2\arctan[(x+y)/\alpha] \approx \pi$ and write
\begin{align}
[\theta_{\gamma}(x), \phi_{\gamma'}(y)] &\approx \delta_{\gamma, \gamma'} \frac{i \pi}{2} \left\{ [\text{sign}(x-y) - 1] + \frac{2 }{\tilde{L}} y \right \} \quad \text{for } x,y \in [0, \tilde{L}]. 
\end{align}

\subsubsection*{Symmetric and antisymmetric fields}
Later on, we will find it useful to work with the symmetric / antisymmetric superpositions of the phase fields
\begin{align}
\vartheta_{\pm}(x) = \frac{1}{\sqrt{2}} [\vartheta_\mathrm{a}(x) \pm \vartheta_\mathrm{b}(x)] \label{EqnApp:def_vartheta_pm}
\end{align}
Using \eq{EqnApp:com_vartheta_ab}, it is straightforward to show that they satisfy the similar relations to $\vartheta_\mathrm{a/b}$, i.e.,
\begin{align}
[\vartheta_{s}(x), \vartheta_{s'}(y)] =& \;\; \delta_{s,s'} i [2\arctan[(x-y)/\alpha] - \pi (x-y) / \tilde{L}] \nonumber \\
\stackrel{\mathclap{\alpha \to 0}}{=} & \;\; \delta_{s,s'} i \pi [\text{sign}(x-y) - (x-y) / \tilde{L}] \quad \text{for } x,y \in [-\tilde{L}, \tilde{L}], \label{EqnApp:com_vartheta_pm}
\end{align}
where $s, s' = \pm$. We also introduce symmetric and antisymmetric combinations of the fields $\theta_\gamma$, $\phi_\gamma$
\begin{align}
\hat \theta_{\pm}(x) &= \frac{1}{\sqrt{2}} [\hat k_\mathrm{a} \pm  \hat k_\mathrm{b} + \theta_\mathrm{a}(x) \pm \theta_\mathrm{b}(x)], \nonumber \\
\phi_{\pm}(x) &= \frac{1}{\sqrt{2}} [\phi_\mathrm{a}(x) \pm \phi_\mathrm{b}(x)],  \label{EqnApp:theta_phi_pm}
\end{align}
where the operators $\hat k_\gamma$ are absorbed by the fields $\hat \theta_{\pm}$. The final bosonized version of our theory will be expressed in terms of these fields. Importantly, the fields from the $+$ and $-$ sector commute as is evident from the previously discussed relations.

\subsection{Bosonization of the Majorana Hamiltonian}
\subsubsection{Free part}
The free part is given by the lattice model
\begin{align}
H_0 =& -t\sum_{\gamma = \mathrm{a,b}}\sum_{j = 1}^{L-1} \left[(c_{\gamma,j}^\dagger c_{\gamma,j+1} + c_{\gamma,j+1}^\dagger c_{\gamma,j}) \right],
\end{align}
which exhibits a dispersion $2 t \cos(k a_0)$ for both species $\gamma = \mathrm{a,b}$ of fermions, where we introduced the lattice constant $a_0$. Assuming a filling fraction $\nu$ such that $k_\mathrm{F} = \frac{\nu  \pi}{a_0}$, this can be seen as a lattice approximation to the continuum theory 
\begin{align}
H_0 &\sim v_\mathrm{F} \sum_\mathrm{\gamma = a. b} \int_0^{\tilde{L}} :\left[\psi_\mathrm{\gamma,R}^\dagger(x) (-i \partial_x) \psi_\mathrm{\gamma,R}(x) + \psi_\mathrm{\gamma,L}^\dagger(x)  (i \partial_x) \psi_\mathrm{\gamma,L}(x) \right]: \mathrm d x \nonumber\\
&= v_\mathrm{F} \sum_\mathrm{\gamma  = a. b} \int_{-\tilde{L}}^{\tilde{L}}  :\left[\psi_\mathrm{\gamma,R}^\dagger(x) (-i \partial_x) \psi_\mathrm{\gamma,R}(x) \right]: \mathrm d x,
\end{align}
where $:...:$ denotes normal-ordering w.r.t. the Fermi surface and $v_\mathrm{F} = 2 t a_0 \sin(k_\mathrm{F} a_0)$. We assume that the continuum fields obey OBC in the sense of \eq{EqnApp:def_OBC} and set $\tilde{L} = (L + 1) a_0$, which can be thought of as adding an additional site at each end of the lattice where the wave functions are zero. \gls{obc} justify the second line as an immediate consequence of \eq{EqnApp:relation_Psi_RL_OBC}. Applying the standard bosonization procedure \cite{Bosonization_Delft} to the $2 \tilde{L}$ periodic fields $\psi_{\gamma,R}(x)$ yields 
\begin{align}
H_0 &\sim v_\mathrm{F}  \sum_\mathrm{\gamma = a. b} \left[\int_{-\tilde{L}}^{\tilde{L}} \frac{1}{2} :(\partial_x \vartheta_\gamma(x) )^2:  \frac{\mathrm d x}{2\pi}  + \frac{\pi}{\tilde{L}} \frac{1}{2} \Delta  N_\gamma (\Delta N_\gamma + 1) \right] \nonumber \\
&= v_\mathrm{F}  \sum_\mathrm{\gamma = a. b} \left[ \sum_{n>0} k_n b_{\gamma,n}^\dagger b_{\gamma,n} + \frac{\pi}{\tilde{L}} \frac{1}{2} \Delta  N_\gamma (\Delta N_\gamma + 1) \right],
\end{align}
where $k_n = n \frac{\pi}{\tilde{L}}$. The finite-size terms do not affect the physics in any relevant way and will vanish in the limit $\tilde{L} \to \infty$, so we neglect them in the following. The remaining part can be expressed through the fields from \eq{EqnApp:def_theta_phi} or \eq{EqnApp:theta_phi_pm} as 
\begin{align}
H_0 \;\; \stackrel{\mathclap{ \tilde{L} \to \infty}}{\sim}& \;\;\; \frac{v_\mathrm{F}}{2 \pi} \sum_\mathrm{\gamma = a, b} \int_{0}^{ \tilde{L}} :\left[ (\partial_x \theta_\gamma(x) )^2 + (\partial_x \phi_\gamma(x) )^2 \right] : \mathrm d x \nonumber \\
=& \;\;\; \frac{v_\mathrm{F}}{2 \pi} \sum_{s = {\pm}} \int_{0}^{\tilde{L}} :\left \{ [\partial_x \hat \theta_s(x)]^2 + [\partial_x \phi_s(x)]^2 \right\} : \mathrm d x.
\end{align}

\subsubsection{Pair hopping}
The pair hopping Hamiltonian is 
\begin{align}
H_W &= W \sum_{j = 1}^{L-1} \left[c_{\mathrm{a},j}^\dagger c_{\mathrm{a},j+1}^\dagger c_{\mathrm{b},j} c_{\mathrm{b},j+1} + \hc \right]
\end{align}
and we start by deriving the associated fermionic continuum theory. To this end, we take the operators on the lattice to be continuum fields evaluated at discrete positions: $c_{\gamma, j} = \sqrt{a_0} \psi_\gamma(j a_0)$. The resulting expression is

\begin{align}
H_W &= W a_0^2  \sum_{j = 1}^{L-1}\left[\psi_\mathrm{a}^\dagger[j a_0] \psi_\mathrm{a}^\dagger[(j+1) a_0] \psi_\mathrm{b}[j a_0] \psi_\mathrm{b}[(j+1) a_0] + \hc \right], \label{EqnApp:pair_hopping}
\end{align}
which we may write in terms of the L/R fields by using \eq{EqnApp:LR_decomposition}. Then, terms with oscillating prefactors $\propto e^{\pm2 i k_\mathrm{F} j a_0}$ and $e^{\pm4i k_\mathrm{F} j a_0}$ appear, whose contributions will integrate out from the long-wavelength effective theory, allowing us to neglect them \cite{Giamarchi}. At filling fraction $\nu = \frac{1}{2}$ corresponding to $k_\mathrm{F} = \frac{\pi}{2 a_0}$, terms with the prefactor $e^{\pm4i k_\mathrm{F} j a_0} = 1$ such as $\psi_\mathrm{a, R}^\dagger [j a_0] \psi_\mathrm{b, L}[j a_0] \psi_\mathrm{a, R}^\dagger[(j+1) a_0] \psi_\mathrm{b, L}[(j+1) a_0]$ should be retained. Based on the bosonization analysis presented later on, we expect these interchain \gls{bs} terms to prevent the formation of a Majorana phase at $\nu = \frac{1}{2}$ in consistency with the numerical results of \cite{Zoller_model}.

Keeping this in mind, we find 
\begin{align}
&\psi_\mathrm{a}^\dagger[j a_0] \psi_\mathrm{a}^\dagger[(j+1) a_0] \psi_\mathrm{b}[j a_0] \psi_\mathrm{b}[(j+1) a_0] \nonumber \\
\sim & \left[\psi_\mathrm{a, R}^\dagger[j a_0] \psi_\mathrm{a, R}^\dagger[(j+1) a_0] \psi_\mathrm{b, R}[j a_0] \psi_\mathrm{b, R}[(j+1) a_0] \right . \nonumber \\
& + \psi_\mathrm{a, R}^\dagger[j a_0] \psi_\mathrm{a, L}^\dagger[(j+1) a_0] \psi_\mathrm{b, R}[j a_0] \psi_\mathrm{b, L}[(j+1) a_0] \nonumber \\
& +  e^{2 i k_\mathrm{F} a_0} \psi_\mathrm{a, R}^\dagger[j a_0] \psi_\mathrm{a, L}^\dagger[(j+1) a_0] \psi_\mathrm{b, L}[j a_0] \psi_\mathrm{b, R}[(j+1) a_0] \nonumber \\
& +  e^{-2 i k_\mathrm{F} a_0} \psi_\mathrm{a, L}^\dagger[j a_0] \psi_\mathrm{a, R}^\dagger[(j+1) a_0] \psi_\mathrm{b, R}[j a_0] \psi_\mathrm{b, L}[(j+1) a_0] \nonumber \\
& + \psi_\mathrm{a, L}^\dagger[j a_0] \psi_\mathrm{a, R}^\dagger[(j+1) a_0] \psi_\mathrm{b, L}[j a_0]\psi_\mathrm{b, R}[(j+1) a_0] \nonumber \\
& \left. + \psi_\mathrm{a, L}^\dagger[j a_0] \psi_\mathrm{a, L}^\dagger [(j+1) a_0]  \psi_\mathrm{b, L}[j a_0]\psi_\mathrm{b, L}[(j+1) a_0] \right].
\end{align}	
This expression contains hopping terms such as $\psi_\mathrm{a, R}^\dagger[j a_0] \psi_\mathrm{a, R}^\dagger[(j+1) a_0] \psi_\mathrm{b, R}[j a_0] \psi_\mathrm{b, R}[(j+1) a_0]$, which will be suppressed by the Pauli principle in the continuum limit. We thus neglect the first and last term of the previous expression. Moving forward, we note that all the fields appearing in the remaining terms such as $\psi_\mathrm{a, R}^\dagger[j a_0] \psi_\mathrm{a, L}^\dagger[(j+1) a_0] \psi_\mathrm{b, R}[j a_0] \psi_\mathrm{b, L}[(j+1) a_0]$ simply anticommute with one another, so the whole expression is already normal-ordered and there is no need for regularization / point splitting later on. Hence, we may neglect the differences of $a_0$ in the spatial arguments of the fields and arrive at the fermionic continuum theory corresponding to \eq{EqnApp:pair_hopping}:

\begin{align}
H_W \sim 2 W a_0 \left[1 - \cos(k_\mathrm{F} a_0) \right] \int_{0}^{\tilde{L}} \left [\psi_\mathrm{a, R}^\dagger(x) \psi_\mathrm{a, L}^\dagger(x) \psi_\mathrm{b, R}(x) \psi_\mathrm{b, L}(x) + \hc\right ] \mathrm d x.
\end{align}

Using \eq{EqnApp:relation_Psi_RL_OBC} and the bosonization identity \eq{EqnApp:Bosonization_ID_2LL} yields
\begin{align}
&\quad\psi_\mathrm{a, R}^\dagger(x) \psi_\mathrm{a, L}^\dagger(x) \psi_\mathrm{b, R}(x) \psi_\mathrm{b, L}(x)  = - \psi_\mathrm{a, R}^\dagger(x) \psi_\mathrm{b, R}(x) \psi_\mathrm{a, R}^\dagger(-x) \psi_\mathrm{b, R}(-x) \nonumber\\
&= - \frac{1}{(2 \pi \alpha)^2} e^{-i \vartheta_\mathrm{a}(x)} e^{-i \frac{\pi}{\tilde{L}} \Delta  N_\mathrm{a} x} e^{-i \hat k_\mathrm{a}}   (-1)^{N_\mathrm{a}}   e^{i \hat k_\mathrm{b}} e^{i \frac{\pi}{\tilde{L}} \Delta  N_\mathrm{b}x} e^{i \vartheta_\mathrm{b}(x)} \nonumber \\
&\quad \times e^{-i \vartheta_\mathrm{a}(-x)} e^{i \frac{\pi}{\tilde{L}} \Delta  N_\mathrm{a}  x} e^{-i \hat k_\mathrm{a}}    (-1)^{N_\mathrm{a}}   e^{i \hat k_\mathrm{b}} e^{-i \frac{\pi}{\tilde{L}} \Delta  N_\mathrm{b}x} e^{i \vartheta_\mathrm{b}(-x)} \nonumber \\
&=e^{-i \frac{\pi}{\tilde{L}} \Delta  N_\mathrm{a} x} e^{i \frac{\pi}{\tilde{L}} (\Delta  N_\mathrm{b}+1) x}   e^{i \frac{\pi}{\tilde{L}} (\Delta  N_\mathrm{a} -1) x}  e^{-i \frac{\pi}{\tilde{L}} (\Delta  N_\mathrm{b}+2) x}  \frac{1}{(2 \pi \alpha)^2} e^{-i \vartheta_\mathrm{a}(x)}  e^{-i \hat k_\mathrm{a}}    e^{i \hat k_\mathrm{b}}  e^{i \vartheta_\mathrm{b}(x)} \nonumber \\
&\quad \times e^{-i \vartheta_\mathrm{a}(-x)} e^{-i \hat k_\mathrm{a}}     e^{i \hat k_\mathrm{b}}  e^{i \vartheta_\mathrm{b}(-x)} \nonumber \\
&= e^{-i 2 \pi x/\tilde{L}} \frac{1}{(2 \pi \alpha)^2} e^{-i [\hat k_\mathrm{a} + \vartheta_\mathrm{a}(x) - \hat k_\mathrm{b}- \vartheta_\mathrm{b}(x)]} e^{-i [\hat k_\mathrm{a} + \vartheta_\mathrm{a}(-x) - \hat k_\mathrm{b}- \vartheta_\mathrm{b}(-x)]} \nonumber \\
&= e^{-i 2 \pi x/\tilde{L}} \frac{1}{(2 \pi \alpha)^2} e^{-i [\hat k_\mathrm{a} - \hat k_\mathrm{b}+ \sqrt{2} \vartheta_- (x)]} e^{-i [\hat k_\mathrm{a} - \hat k_\mathrm{b}+ \sqrt{2} \vartheta_-(-x) ]}.
\end{align}
Remember that the operators $\hat k_\gamma$ commute with the fields $\vartheta$. If $[X, [X,Y]] = 0$ and $[Y, [X,Y]] = 0$, the \gls{bch} identity $e^{X} e^{Y} = e^{X+Y} e^{[X,Y]/2}$ holds, and we may use the commutator \eq{EqnApp:com_vartheta_pm} to write
\begin{align}
e^{-i \sqrt{2} \vartheta_- (x)} e^{-i \sqrt{2} \vartheta_-(-x)} &= e^{-i \sqrt{2} [\vartheta_- (x) + \vartheta_- (-x)]} e^{-[\vartheta_- (x), \vartheta_- (-x)]} \nonumber \\
&= e^{-i \sqrt{2} [\vartheta_- (x) + \vartheta_- (-x)]} e^{-i\pi[ 1 - 2x/\tilde{L}} = -e^{-i \sqrt{2} [\vartheta_- (x) + \vartheta_- (-x)]} e^{i 2 \pi x/\tilde{L}}
\end{align}
The phase factor $e^{i 2 \pi x/\tilde{L}}$ precisely cancels the one from the penultimate equation. We arrive at
\begin{align}
\psi_\mathrm{a, R}^\dagger(x) \psi_\mathrm{a, L}^\dagger(x) \psi_\mathrm{b, R}(x) \psi_\mathrm{b, L}(x)  = -\frac{1}{(2 \pi \alpha)^2} e^{-i \sqrt{2} [\sqrt{2} (\hat k_\mathrm{a} - \hat k_\mathrm{b}) + \vartheta_- (x) + \vartheta_- (-x)]}
\end{align}
Using \eq{EqnApp:def_theta_phi}, \eq{EqnApp:def_vartheta_pm}, and \eq{EqnApp:theta_phi_pm} leads to the final expression 
\begin{align}
H_W \sim \frac{4[\cos(2 k_\mathrm{F} a_0) - 1] W a_0}{(2 \pi \alpha)^2} \int_0^{\tilde{L}} \cos(\sqrt{8} \hat \theta_-) \mathrm d x. 
\end{align}

\subsection{Bosonization of flux hopping}
The lattice Hamiltonian 
\begin{align}
H_\phi &= r \sum_{j = 1}^{L}  \left[ e^{2 \pi i \phi j}c_{\mathrm{a},j}^\dagger c_{\mathrm{b},j} + e^{-2 \pi i \phi j}c_{\mathrm{b},j}^\dagger c_{\mathrm{a},j} \right], \label{EqnApp:H_phi}
\end{align}
is mapped to a continuum fermionic theory in the same way as before:
\begin{align}
H_\phi &\sim r \int_{0}^{\tilde{L}} \left\{ e^{2 \pi i \phi x/a_0} \left[ \psi_\mathrm{R, a}^\dagger(x) \psi_\mathrm{R, b}(x) + \psi_\mathrm{L, a}^\dagger(x) \psi_\mathrm{L, b}(x) + e^{-2i k_\mathrm{F} x} \psi_\mathrm{R, a}^\dagger(x) \psi_\mathrm{L, b} [x]+  e^{2 i k_\mathrm{F} x} \psi_\mathrm{L, a}^\dagger(x) \psi_\mathrm{R, b}(x)\right] + \hc \right \}\mathrm d x. \label{EqnApp:H_phi_continuum} 
\end{align}
In general, the oscillating prefactors will suppress this term from the effective low-energy theory. At $\phi = 0$, the \gls{fs} terms survive, while at $\phi = \pm \nu$, one of the \gls{bs} terms will make an impact because $2 \pi \phi$ exactly cancels $2k_\mathrm{F} = \frac{2 \pi \nu }{a_0}$.

\subsubsection{Bosonization for $\phi = 0$}
At $\phi = 0$, \eq{EqnApp:H_phi_continuum} reduces to regular \gls{fs}. We apply the bosonization identity \eq{EqnApp:Bosonization_ID_2LL} to find 
\begin{align}
\left[\psi_\mathrm{R, a}^\dagger(x) \psi_\mathrm{R, b}(x) + \hc \right] &= \frac{1}{2 \pi \alpha} \left [e^{-i \vartheta_\mathrm{a}(x)} e^{-i \frac{\pi}{\tilde{L}} \Delta  N_\mathrm{a} x} e^{-i \hat k_\mathrm{a}}   (-1)^{N_\mathrm{a}}  e^{i \frac{\pi}{\tilde{L}} (\Delta  N_\mathrm{b} + 1) x} e^{i \hat k_\mathrm{b}} e^{i \vartheta_\mathrm{b}(x)} \right . \nonumber \\
&\quad+ \left .  e^{- i \vartheta_\mathrm{b}(x)} e^{-i \frac{\pi}{\tilde{L}} \Delta N_\mathrm{b} x} e^{- i \hat k_\mathrm{b}}(-1)^{N_\mathrm{a}} e^{i \frac{\pi}{\tilde{L}} (\Delta  N_\mathrm{a} + 1) x} e^{i \hat k_\mathrm{a}} e^{i \vartheta_\mathrm{a}(x)} \right] \nonumber \\
&= \frac{1}{2 \pi \alpha} (-1)^{N_\mathrm{a}} \left[e^{i \frac{\pi}{\tilde{L}} (N_\mathrm{b}  - N_\mathrm{a} + 1) x} e^{i[\hat k_\mathrm{a} - \hat k_\mathrm{b} + \sqrt{2} \vartheta_-(x)]} - \hc\right]
\end{align}
and similarly 
\begin{align}
\left[\psi_\mathrm{L, a}^\dagger(x) \psi_\mathrm{L, b}(x) + \hc \right] = \left[\psi_\mathrm{R, a}^\dagger(-x) \psi_\mathrm{R, b}(-x) + \hc \right]  = \frac{1}{2 \pi \alpha} (-1)^{N_\mathrm{a}} \left[e^{-i \frac{\pi}{\tilde{L}} (N_\mathrm{b}  - N_\mathrm{a} + 1) x} e^{i[\hat k_\mathrm{a} - \hat k_\mathrm{b} + \sqrt{2} \vartheta_-(-x)]} - \hc\right].
\end{align}
Given the relations between the various fields, it is readily seen that 
\begin{align}
\hat k_\mathrm{a} - \hat k_\mathrm{b} +  \sqrt{2} \vartheta_-(\pm x) = \sqrt{2} \hat \theta_-(x) \pm \sqrt{2} \phi_-(x),
\end{align}
which leads to the bosonized expression stated in the main text
\begin{align}
H_{\phi = 0} &\sim r \frac{(-1)^{N_\mathrm{a}}}{2 \pi \alpha} \int_{0}^{\tilde{L}}  \left[e^{i \frac{\pi}{\tilde{L}} (N_\mathrm{b}  - N_\mathrm{a} + 1) x} e^{i\sqrt{2}  [\hat \theta_-(x) + \phi_-(x)]} +  e^{- i \frac{\pi}{\tilde{L}} (N_\mathrm{b}  -  N_\mathrm{a} + 1) x} e^{i\sqrt{2} [\hat \theta_-(x) - \phi_-(x)]} - \hc \right] \mathrm d x.
\end{align}

In the main text, we argue that the relative minus sign arising from the finite size term $\pi (N_\mathrm{b}  - N_\mathrm{a} + 1) x / (\tilde{L})$ will cancel contributions to the \gls{gs} splitting from the left and right end. While this is more of a qualitative argument, the cancellation can be shown on an exact level by considering that in addition to parity symmetry, the Hamiltonian $H_0 + H_W$ is also invariant under the action of the unitary inversion symmetry $U_\mathrm{I}$ defined by
\begin{align}
U_\mathrm{I} c_{\gamma, j} U_\mathrm{I}^\dagger = c_{\gamma, N - j + 1}, \quad U_\mathrm{I} c^\dagger_{\gamma, j} U_\mathrm{I}^\dagger = c^\dagger _{\gamma, N - j + 1}. \label{EqnApp:U_I}
\end{align}
This is nothing but a mirroring at the center of the chain. $U_\mathrm{I}$ squares to one, which restricts the possible eigenvalues to $\pm 1$,  and commutes with $P_\mathrm{a}$, so the two \gls{gs} $\ket{P_\mathrm{a} = \pm 1}$ are also eigenstates of $U_\mathrm{I}$. 
To determine the relative contribution of fluxless hoppings to the \gls{gs} splitting from the left and right end, we consider a collection of hoppings located on the left side of the chain and denote it by $H_{\phi = 0}^\mathrm{L}$. Inversion symmetry maps this to the mirrored set of hoppings on the opposite side: $U_\mathrm{I} H_{\phi = 0}^\mathrm{L} U_\mathrm{I}^\dagger = H_{\phi = 0}^\mathrm{R}$. At small enough hopping amplitude $r$, it is sufficient to take into account the matrix elements between the two \gls{gs} to determine the splitting as the antisymmetric sector exhibits a large excitation gap. Since the operators will change the parity, we only need to look at the matrix element

\begin{align}
\bra{P_\mathrm{a} = 1} H_{\phi = 0}^\mathrm{R} \ket{P_\mathrm{a} = -1} = \bra{P_\mathrm{a} = 1} U_\mathrm{I} H_{\phi = 0}^\mathrm{L} U_\mathrm{I}^\dagger \ket{P_\mathrm{a} = -1} = u_\mathrm{I,+} u_\mathrm{I,-}\bra{P_\mathrm{a} = 1} H_{\phi = 0}^\mathrm{L} \ket{P_\mathrm{a} = -1}.
\end{align}
Here, $u_\mathrm{I,\pm}$ denotes the eigenvalue of $U_\mathrm{I}$ associated with the positive or negative parity eigenstate $U_\mathrm{I} \ket{P_\mathrm{a} = \pm 1} =  u_\mathrm{I,\pm} \ket{P_\mathrm{a} = \pm 1}$. Depending on whether these eigenvalues have the same or opposite signs, the contributions from the left and right end will either amplify or cancel exactly. The relative sign that we derive from the field-theoretical analysis suggests that $u_\mathrm{I,\pm}$ will have opposite sign for even $N_\mathrm{tot}$ and same sign for odd $N_\mathrm{tot}$ in consistency with numerical data. 

\subsubsection{Bosonization for $\phi = \nu$}
We derive the bosonization of \eq{EqnApp:H_phi} for $\phi = \nu$ here, the case of $\phi = -\nu$ can be treated on similar footing. Applying \eq{EqnApp:Bosonization_ID_2LL} to the \gls{bs} term appearing in \eq{EqnApp:H_phi_continuum} yields
\begin{align}
&\quad[\psi_\mathrm{R, a}^\dagger(x) \psi_\mathrm{L, b}(x) + \psi_\mathrm{L, b}^\dagger(x) \psi_\mathrm{R, a}(x) + \hc] = -[\psi_\mathrm{R, a}^\dagger(x) \psi_\mathrm{R, b}(-x) + \psi_\mathrm{R, b}^\dagger(-x) \psi_\mathrm{R, a}(x) + \hc] \nonumber \\
&= - \left[ e^{-i \vartheta_\mathrm{a}(x)} e^{-i \frac{\pi}{\tilde{L}} \Delta  N_\mathrm{a} x} e^{-i \hat k_\mathrm{a}} \frac{(-1)^{N_\mathrm{a}}}{2 \pi \alpha} e^{-i \frac{\pi}{\tilde{L}} (\Delta  N_\mathrm{b} + 1)x} e^{i \hat k_\mathrm{b}}  e^{i \vartheta_\mathrm{b}(-x)} + \hc \right ] \nonumber \\
&=\frac{(-1)^{N_\mathrm{a}}}{2 \pi \alpha} \left[ e^{-i \frac{\pi}{\tilde{L}} (\Delta  N_\mathrm{a} + \Delta  N_\mathrm{b} + 1) x}  e^{-i[\hat k_\mathrm{a} - \hat k_\mathrm{b} +  \vartheta_\mathrm{a}(x) -  \vartheta_\mathrm{b}(-x)]} - \hc \right ] 
\end{align}
We have $\vartheta_\mathrm{a}(x) - \vartheta_\mathrm{b}(-x) = [\phi_\mathrm{a}(x) + \phi_\mathrm{b}(x) + \theta_\mathrm{a}(x) - \theta_\mathrm{b}(x)]$. Keeping in mind that the fields from the symmetric and antisymmetric sector commute and that all operators without hat commute with the particle numbers, we write 
\begin{align}
[\psi_\mathrm{R, a}^\dagger(x) \psi_\mathrm{L, b}(x) + \psi_\mathrm{L, b}^\dagger(x) \psi_\mathrm{R, a}(x) + \hc] =\frac{(-1)^{N_\mathrm{a}}}{2 \pi \alpha} \left[ e^{-i [\frac{\pi}{\tilde{L}} (\Delta  N_\mathrm{a} + \Delta  N_\mathrm{b} + 1) x + \sqrt{2} \phi_+(x)]}  e^{-i \sqrt{2} \hat \theta_-(x)} - \hc \right ]. 
\end{align}
Because $e^{\pm i (k_\mathrm{a} - k_\mathrm{b})}$ does not change the total particle number, $\Delta  N_\mathrm{a} + \Delta  N_\mathrm{b}$ commutes with $e^{\pm i \sqrt{2} \hat \theta_-(x)}$, allowing us to write
\begin{align}
H_{\phi = \nu} &\sim \frac{-i r (-1)^{N_\mathrm{a}}}{\pi \alpha}  \int_{0}^{\tilde{L}} \left \{ \sin \left [\frac{\pi}{\tilde{L}} (\Delta  N_\mathrm{a} + \Delta  N_\mathrm{b} + 1) x + \sqrt{2} \phi_+(x) \right] \cos \left[ \sqrt{2} \hat \theta_-(x) \right]  \right. \nonumber \\
&\quad + \left . \cos \left [\frac{\pi}{\tilde{L}} (\Delta  N_\mathrm{a} + \Delta  N_\mathrm{b} + 1) x + \sqrt{2} \phi_+(x) \right] \sin \left[ \sqrt{2} \hat \theta_-(x) \right]  \right \} \mathrm d x. 
\end{align}
We have $k_\mathrm{F} = \frac{\nu  \pi}{a_0}$ and $\tilde{L} = (L+1) a_0$, so the finite-size term is $\Delta N_\mathrm{a} + \Delta N_\mathrm{b} = N_\mathrm{a} + N_\mathrm{b} - 2 k_F \tilde{L} / \pi = N_\mathrm{tot} - 2 \nu (L + 1)$, which yields the expression stated in the main text. 
	
To conclude this section, we argue why $H_{\phi = \nu}$ will always lift the finite-size gap based on a mean-field treatment. The bosonized version of the base model decouples into a symmetric and an antisymmetric sector, i.e., $H_0 + H_W \sim H_+ \otimes \mathbb I_- +  \mathbb I_+\otimes H_- $, such that the eigenstates can be thought of as tensor products $\ket{\psi}_+ \otimes \ket{\psi'}_-$ of $H_+$ and $H_-$ eigenstates. In the antisymmetric sector, there are two degenerate \gls{gs}  $\ket{\theta_1}_-$ and $ \ket{\theta_2}_-$ separated from the rest by a large gap, allowing us to restrict the antisymmetric sector to these two states in the spirit of degenerate perturbation theory. At negative $W$, the (Hermitian!) term $i (-1)^{N_\mathrm{a}} \cos \left[ \sqrt{2} \hat \theta_-(x) \right] $ is zero in this restricted subspace, while $i (-1)^{N_\mathrm{a}} \sin \left[ \sqrt{2} \hat \theta_-(x) \right] $ has some non-trivial action ($\sin \left[ \sqrt{2} \hat \theta_-(x) \right]$ yields $\pm 1$ when applied to the states $\ket{\theta_1}_-$, $\ket{\theta_2}_-$, while $(-1)^{N_\mathrm{a}}$ exchanges them). After diagonalizing this operator in the two-state space, the Hilbert space further decouples into states of the form $\ket{\psi}_+ \otimes \ket{\Gamma_+}_-$ and $\ket{\psi}_+ \otimes \ket{\Gamma_-}_-$, where $\Gamma_{\pm}$ denotes the eigenvalue of $i (-1)^{N_\mathrm{a}} \cos \left[ \sqrt{2} \hat \theta_-(x) \right] $.
In these two subspaces, we are left with the Sine-Gordon theory

\begin{align}
\frac{v_\mathrm{F}}{2 \pi} \int_{0}^{\tilde{L}} \left \{ [\partial_x \hat \theta_+(x)]^2 + [\partial_x \phi_+(x)]^2 \right\}  \mathrm d x -  \frac{r \Gamma_{\pm} }{\pi \alpha}  \int_{0}^{\tilde{L}} \cos \left [ \sqrt{2} \phi_+(x) \right] \mathrm d x,
\end{align}
for which standard RG-flow equations indicate the formation of a massive phase for any value of $r$ \cite{Giamarchi}, in consistency with numerical data. The same line of reasoning applies to the case $W> 0$.

\section{Critical state preparation with a spatially inhomogeneous ramp} \label{App:Sec:IH_ramp}
Spatially inhomogeneous ramps of the mass term have also been proposed in the literature as a way to achieve optimal (i.e., $\propto 1/L$) scaling of the preparation time $\tau_\mathrm{tot}$ \cite{crit_state_prep_3, crit_state_prep_4}. However, in the present case, we find that the homogeneous ramp with a power $p$ adjusted to system size significantly outperforms the inhomogeneous ramp approach. Concretely, we implement a procedure oriented on Ref.~\cite{crit_state_prep_4}, but with the generalization to multiple critical fronts that propagate in space with a velocity $v_\mathrm{r}$. For this, we introduce the ramp function
\begin{align}
\epsilon(u) = \mathbbm{1}_{(-\infty, \pi/2]}(u) + 0.5 [1 - \sin(u)] \mathbbm{1}_{(-\pi/2, \pi/2]}(u), \label{EqnApp:def_epsilon}
\end{align}
and the auxiliary function 
\begin{align}
u(x, \tau) = \min_{l = 1, ..., n_\mathrm{r}}\alpha\left[v \tau  - |x - x_l| - d/2 \right], \label{EqnApp:def_epsilon}
\end{align}
where $x_l = 1 + \Delta_{n_\mathrm{r}} (2l - 1) $ is the starting position of the $l$th front, the spacing is $\Delta_{n_\mathrm{r}} = (L - 1) / (2 n_r)$, and the offset is $d = \pi / \alpha$. The composition $\epsilon(u(x, \tau))$ describes $n_\mathrm{r}$ fronts starting at evenly spaced points on the interval $[1, L]$ and propagating at velocity $v$ through the system. The parameter $\alpha$ controls the smoothness of the ramp by smearing out each front over the distance $d$. 

We then consider the flux-hopping Hamiltonian with spatially varying and time-dependent amplitudes  
\begin{align}
H_\phi(\tau) &=  \sum_{j = 1}^L  r_j(\tau) \left[ e^{2 \pi i \phi j}c_{\mathrm{a},j}^\dagger c_{\mathrm{b},j} + e^{-2 \pi i \phi j}c_{\mathrm{b},j}^\dagger c_{\mathrm{a},j} \right] \label{EqnApp:H_phi_r_j}
\end{align}
and set their time-dependence to $r_j(\tau) = r(j, \tau)$ by introducing the function
\begin{align}
r(x,\tau) = r_0 \epsilon(u(x, \tau)). \label{EqnApp:def_r_x_t}
\end{align}
Other than that, we still work at an exact filling fraction of $\nu = 1/3$ and set $\phi = \nu = 1/3$. At $\tau = 0$, all couplings are set to $r_j  = r_0$, which we put to $r_0 = 0.1$, thereby starting at the beginning of stage three of \eq{Eqn:time_dependence_p1}. We illustrate the ramp function at two different times in Fig.~\ref{Fig:fidelity_data_appendix_a} for $n_\mathrm{r} = 2$ fronts, velocity $v_{n_\mathrm{r} = 2} = 0.1$, smoothness parameter $\alpha = 1/4$, and a system size of $L = 48$. The relation between the time $\tau_\mathrm{tot}$ to complete the protocol in the sense of arriving at $r_j(\tau_\mathrm{tot}) = 0 \; \forall j$ and the ramp velocity $v_\mathrm{r}$ is
\begin{align}
\tau_\mathrm{tot} = \left[d + \frac{L-1}{2 n_\mathrm{r}} \right] \frac{1}{v_{n_\mathrm{r}}}\; \Leftrightarrow \; v_{n_\mathrm{r}} = \left[d + \frac{L - 1}{2 n_\mathrm{r}} \right] \frac{1}{\tau_\mathrm{tot}}, \label{EqnApp:t_tot_v_r}
\end{align}
where the constant offset is due to the finite width $d = \pi / \alpha$ of the critical front. 

We conduct \gls{mpste} simulations to compare this approach against the strategy of a global ramp with a power law $p(L)$ adjusted to system size that we present in the main text. In general, we find that a smoothness $\alpha \lesssim 1/4$ is sufficient for adiabatic preparation. However, contrary to the claims of \cite{crit_state_prep_3}, the ramp speed $v_\mathrm{r}$ has to be adapted to system size to keep the time evolution adiabatic when starting only a single front, corresponding to $n_\mathrm{r} = 1$. While we find that this can be countered to some degree by starting multiple fronts for larger systems, the strategies presented in the main text outperform this procedure in either case for the investigated system sizes, with a clear trend of the advantage to increase with system size. Concretely, we present data for system sizes $L = 24$, $L = 48$, and $L = 72$ in Fig.~\ref{Fig:fidelity_data_appendix_b} to Fig.~\ref{Fig:fidelity_data_appendix_d} for the case of $n_\mathrm{r} = 1$ (similar to the study in \cite{crit_state_prep_3}) and an increasing value of $n_\mathrm{r}$.

\begin{figure}[htp!]	 
{
        \vbox to 0pt {
                \raggedright
                \textcolor{white}{
                    \subfloatlabel[1][Fig:fidelity_data_appendix_a]
                    \subfloatlabel[2][Fig:fidelity_data_appendix_b]
                    \subfloatlabel[3][Fig:fidelity_data_appendix_c]
                    \subfloatlabel[4][Fig:fidelity_data_appendix_d]
                }
            }
    }
    \includegraphics[width=\textwidth]{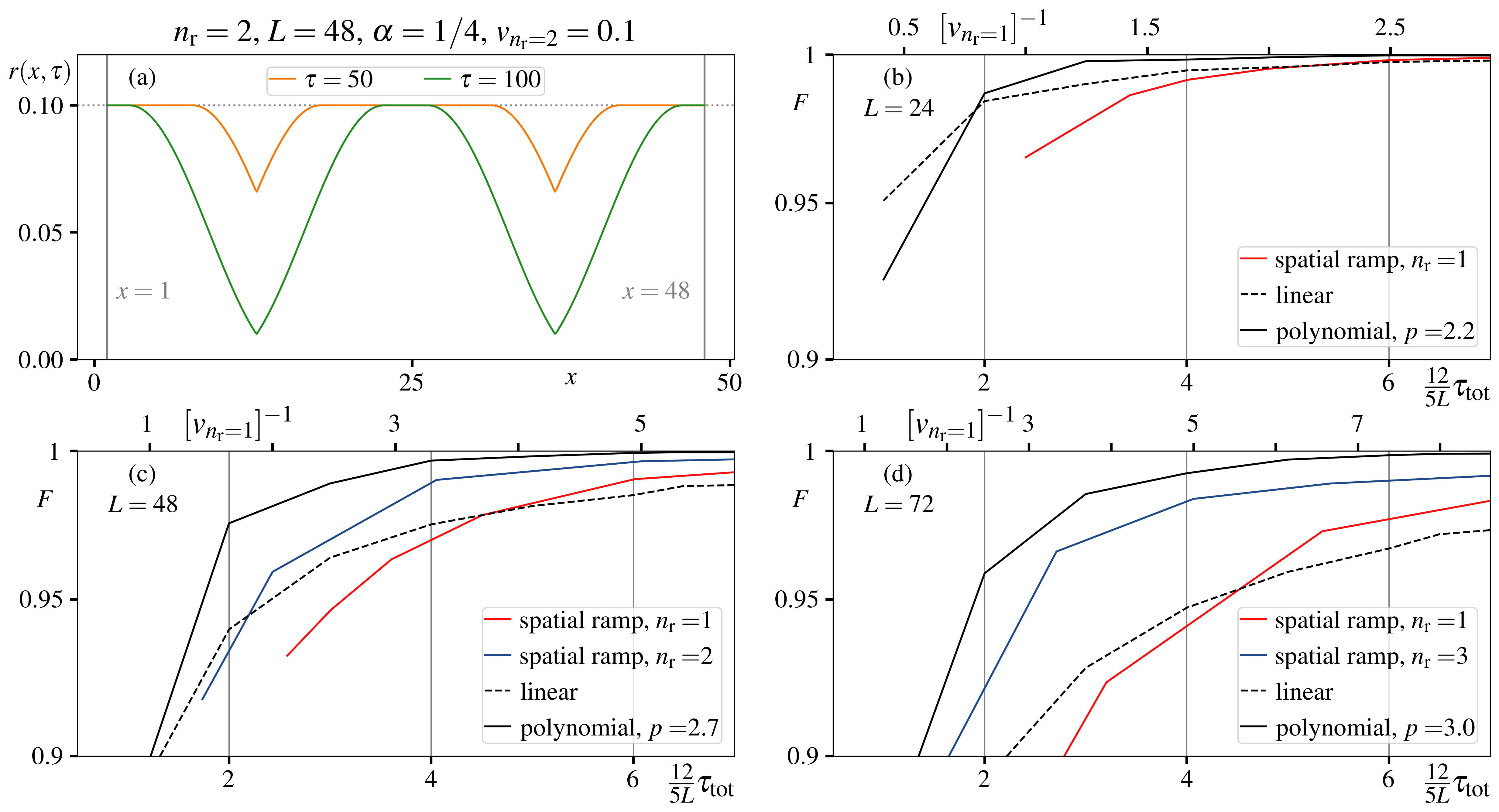}
\caption{(a) Illustration of the ramp function $r(x, \tau)$ as per \eq{EqnApp:def_r_x_t} for a system size of $L = 48$, with $r_0 = 0.1$, $v_{n_\mathrm{r} = 2} = 0.1$, $\alpha  =1/4$, and $n_\mathrm{r} = 2$ at times $\tau= 50$ and $\tau = 100$. (b) \gls{gs} fidelity $F$ after adiabatic evolution with the inhomogeneous ramp for $n_\mathrm{r} = 1$, $\alpha = 1/4$ as a function of preparation time $\tau_\mathrm{tot}$ for a system of size $L = 24$ at filling  $\nu = 1/3$ with the time axis rescaled proportional to system size by $12 / (5 L)$. Other Parameters are $t = 1$, $W = -1.8$, and $\phi = 1/3$. Additionally, the reciprocal $[v_{n_\mathrm{r} = 1}]^{-1}$ of the corresponding ramp velocity as per \eq{EqnApp:t_tot_v_r} is indicated on the upper axis. For comparison, the fidelity curves of the global ramp protocol discussed in the main text (cf. Fig.~(\ref{Fig:fidelity_data})) are also shown in black. (c) Similar data for a system size of $L = 48$ at filling $\nu = 1/3$. We present data for a single front $n_\mathrm{r} = 1$ (red line) and two fronts $n_\mathrm{r} = 2$ (blue line) in comparison to the data from the global ramp. The reciprocal velocity associated to the $n_\mathrm{r} = 1$ case is again indicated on the upper axis. (d) Similar to (c), but for $L = 72$ at filling $\nu = 1/3$ and with the blue line representing the fidelity for $n_\mathrm{r} = 3$ critical fronts.}
\end{figure}

\end{document}